\newcommand*\patchAmsMathEnvironmentForLineno[1]{%
  \expandafter\let\csname old#1\expandafter\endcsname\csname #1\endcsname
  \expandafter\let\csname oldend#1\expandafter\endcsname\csname end#1\endcsname
  \renewenvironment{#1}%
     {\linenomath\csname old#1\endcsname}%
     {\csname oldend#1\endcsname\endlinenomath}}%
\newcommand*\patchBothAmsMathEnvironmentsForLineno[1]{%
  \patchAmsMathEnvironmentForLineno{#1}%
  \patchAmsMathEnvironmentForLineno{#1*}}%
\AtBeginDocument{%
\patchBothAmsMathEnvironmentsForLineno{equation}%
\patchBothAmsMathEnvironmentsForLineno{align}%
}
\patchAmsMathEnvironmentForLineno{equation}

\documentclass[aps,prd,nofootinbib,twocolumn,floatfix,superscriptaddress]{revtex4-1}

\usepackage[mathlines]{lineno}
\usepackage{natbib}			   
\usepackage{amsmath}                
\usepackage{amssymb}                
\usepackage{graphicx}
\usepackage{verbatim}              
\usepackage{fancyhdr}
\usepackage{bm}

\usepackage{dcolumn}		

\usepackage{longtable}
\usepackage{tabularx, blindtext} 

\usepackage{ifpdf}
\ifpdf
\DeclareGraphicsExtensions{.pdf, .jpg, .tif}
\usepackage[
      pdftex,                  
      colorlinks=true,          
      urlcolor=blue,           
      menucolor=black,          
      linkcolor=blue, 
      anchorcolor=blue,  
      citecolor=blue,
      filecolor=blue,
      bookmarks=true,           
      bookmarksopen=true,       
      hyperfootnotes=false,     
      pdfpagemode=UseOutlines,  
      pdfauthor={The HERMES Collaboration},
      pdftitle={Kinematic Dependence of 
	Longitudinal Double-Spin Asymmetries at HERMES},
]{hyperref}
\else
\DeclareGraphicsExtensions{.eps, .jpg}
\usepackage[%
  breaklinks=true,%
  colorlinks=true,%
  linkcolor=blue,anchorcolor=blue,%
  citecolor=blue,filecolor=blue,%
  menucolor=blue,pagecolor=blue,%
  urlcolor=blue]{hyperref}
\fi

\allowdisplaybreaks 
\fancypagestyle{titlepage}{\fancyhead[R]{\version ~-- \today}}

\newcommand{\de}{{\rm\,d}}

  \newcommand{\approptoinn}[2]{\mathrel{\vcenter{
  \offinterlineskip\halign{\hfil$##$\cr
    #1\propto\cr\noalign{\kern2pt}#1\sim\cr\noalign{\kern-2pt}}}}}

\newcommand{\version}{DESY 18-181  --  v4.0}

\begin{document}

\title{
  Longitudinal double-spin asymmetries in semi-inclusive deep-inelastic scattering of electrons and positrons by protons and deuterons
}


\def\groupargonne{\affiliation{Physics Division, Argonne National Laboratory, Argonne, Illinois 60439-4843, USA}}
\def\groupbari{\affiliation{Istituto Nazionale di Fisica Nucleare, Sezione di Bari, 70124 Bari, Italy}}
\def\groupbeijing{\affiliation{School of Physics, Peking University, Beijing 100871, China}}
\def\groupbilbao{\affiliation{Department of Theoretical Physics, University of the Basque Country UPV/EHU, 48080 Bilbao, Spain and IKERBASQUE, Basque Foundation for Science, 48013 Bilbao, Spain}}
\def\groupcolorado{\affiliation{Nuclear Physics Laboratory, University of Colorado, Boulder, Colorado 80309-0390, USA}}
\def\groupdesy{\affiliation{DESY, 22603 Hamburg, Germany}}
\def\groupzeuthen{\affiliation{DESY, 15738 Zeuthen, Germany}}
\def\groupdubna{\affiliation{Joint Institute for Nuclear Research, 141980 Dubna, Russia}}
\def\grouperlangen{\affiliation{Physikalisches Institut, Universit\"at Erlangen-N\"urnberg, 91058 Erlangen, Germany}}
\def\groupferrara{\affiliation{Istituto Nazionale di Fisica Nucleare, Sezione di Ferrara and Dipartimento di Fisica e Scienze della Terra, Universit\`a di Ferrara, 44122 Ferrara, Italy}}
\def\groupfrascati{\affiliation{Istituto Nazionale di Fisica Nucleare, Laboratori Nazionali di Frascati, 00044 Frascati, Italy}}
\def\groupgent{\affiliation{Department of Physics and Astronomy, Ghent University, 9000 Gent, Belgium}}
\def\groupgiessen{\affiliation{II. Physikalisches Institut, Justus-Liebig Universit\"at Gie{\ss}en, 35392 Gie{\ss}en, Germany}}
\def\groupglasgow{\affiliation{SUPA, School of Physics and Astronomy, University of Glasgow, Glasgow G12 8QQ, United Kingdom}}
\def\groupillinois{\affiliation{Department of Physics, University of Illinois, Urbana, Illinois 61801-3080, USA}}
\def\groupmichigan{\affiliation{Randall Laboratory of Physics, University of Michigan, Ann Arbor, Michigan 48109-1040, USA }}
\def\groupmoscow{\affiliation{Lebedev Physical Institute, 117924 Moscow, Russia}}
\def\groupnikhef{\affiliation{National Institute for Subatomic Physics (Nikhef), 1009 DB Amsterdam, The Netherlands}}
\def\groupstpetersburg{\affiliation{B.P. Konstantinov Petersburg Nuclear Physics Institute, Gatchina, 188300 Leningrad Region, Russia}}
\def\groupprotvino{\affiliation{Institute for High Energy Physics, Protvino, 142281 Moscow Region, Russia}}
\def\groupregensburg{\affiliation{Institut f\"ur Theoretische Physik, Universit\"at Regensburg, 93040 Regensburg, Germany}}
\def\grouprome{\affiliation{Istituto Nazionale di Fisica Nucleare, Sezione di Roma, Gruppo Collegato Sanit\`a and Istituto Superiore di Sanit\`a, 00161 Roma, Italy}}
\def\grouptriumf{\affiliation{TRIUMF, Vancouver, British Columbia V6T 2A3, Canada}}
\def\grouptokyo{\affiliation{Department of Physics, Tokyo Institute of Technology, Tokyo 152, Japan}}
\def\groupamsterdam{\affiliation{Department of Physics and Astronomy, VU University, 1081 HV Amsterdam, The Netherlands}}
\def\groupwarsaw{\affiliation{National Centre for Nuclear Research, 00-689 Warsaw, Poland}}
\def\groupyerevan{\affiliation{Yerevan Physics Institute, 375036 Yerevan, Armenia}}
\def\groupnone{\noaffiliation}


\groupargonne
\groupbari
\groupbeijing
\groupbilbao
\groupcolorado
\groupdesy
\groupzeuthen
\groupdubna
\grouperlangen
\groupferrara
\groupfrascati
\groupgent
\groupgiessen
\groupglasgow
\groupillinois
\groupmichigan
\groupmoscow
\groupnikhef
\groupstpetersburg
\groupprotvino
\grouprome
\grouptriumf
\grouptokyo
\groupamsterdam
\groupwarsaw
\groupyerevan


\author{A.~Airapetian}  \groupgiessen \groupmichigan
\author{N.~Akopov}  \groupyerevan
\author{Z.~Akopov}  \groupdesy
\author{E.C.~Aschenauer}  \groupzeuthen
\author{W.~Augustyniak}  \groupwarsaw
\author{R.~Avakian}  \groupyerevan
\author{A.~Avetissian}  \groupyerevan
\author{S.~Belostotski}  \groupstpetersburg
\author{H.P.~Blok}  \groupnikhef \groupamsterdam
\author{A.~Borissov}  \groupdesy
\author{V.~Bryzgalov}  \groupprotvino
\author{G.P.~Capitani}  \groupfrascati
\author{E.~Cisbani}  \grouprome
\author{G.~Ciullo}  \groupferrara
\author{M.~Contalbrigo}  \groupferrara
\author{P.F.~Dalpiaz}  \groupferrara
\author{W.~Deconinck}  \groupdesy
\author{R.~De~Leo}  \groupbari
\author{L.~De~Nardo}  \groupdesy \groupgent \grouptriumf
\author{E.~De~Sanctis}  \groupfrascati
\author{M.~Diefenthaler}  \grouperlangen 
\author{P.~Di~Nezza}  \groupfrascati
\author{M.~D\"uren}  \groupgiessen
\author{G.~Elbakian}  \groupyerevan
\author{F.~Ellinghaus}  \groupcolorado
\author{A.~Fantoni}  \groupfrascati
\author{L.~Felawka}  \grouptriumf
\author{S.~Frullani}\thanks{Deceased}  \grouprome 
\author{G.~Gavrilov}  \groupdesy \groupstpetersburg \grouptriumf
\author{V.~Gharibyan}  \groupyerevan
\author{F.~Giordano}  \groupferrara 
\author{S.~Gliske}  \groupmichigan
\author{D.~Hasch}  \groupfrascati
\author{Y.~Holler}  \groupdesy
\author{A.~Ivanilov}  \groupprotvino
\author{H.E.~Jackson}  \groupargonne
\author{S.~Joosten}  \groupgent 
\author{R.~Kaiser}  \groupglasgow
\author{G.~Karyan}  \groupyerevan
\author{T.~Keri}  \groupgiessen \groupglasgow
\author{E.~Kinney}  \groupcolorado
\author{A.~Kisselev}  \groupstpetersburg
\author{V.~Korotkov}\thanks{Deceased}  \groupprotvino
\author{V.~Kozlov}  \groupmoscow
\author{P.~Kravchenko}  \grouperlangen \groupstpetersburg
\author{V.G.~Krivokhijine}  \groupdubna
\author{L.~Lagamba}  \groupbari
\author{L.~Lapik\'as}  \groupnikhef
\author{I.~Lehmann}  \groupglasgow
\author{W.~Lorenzon}  \groupmichigan
\author{B.-Q.~Ma}  \groupbeijing
\author{D.~Mahon}  \groupglasgow
\author{S.I.~Manaenkov}  \groupstpetersburg
\author{Y.~Mao}  \groupbeijing
\author{B.~Marianski}  \groupwarsaw
\author{H.~Marukyan}  \groupyerevan
\author{Y.~Miyachi}  \grouptokyo
\author{A.~Movsisyan}  \groupferrara \groupyerevan
\author{V.~Muccifora}  \groupfrascati
\author{A.~Mussgiller}  \groupdesy \grouperlangen
\author{Y.~Naryshkin}  \groupstpetersburg
\author{A.~Nass}  \grouperlangen
\author{G.~Nazaryan}  \groupyerevan
\author{W.-D.~Nowak}  \groupzeuthen
\author{L.L.~Pappalardo}  \groupferrara
\author{R.~Perez-Benito}  \groupgiessen
\author{A.~Petrosyan}  \groupyerevan
\author{P.E.~Reimer}  \groupargonne
\author{A.R.~Reolon}  \groupfrascati
\author{C.~Riedl}  \groupzeuthen \groupillinois
\author{K.~Rith}  \grouperlangen
\author{G.~Rosner}  \groupglasgow
\author{A.~Rostomyan}  \groupdesy
\author{J.~Rubin}  \groupillinois 
\author{D.~Ryckbosch}  \groupgent
\author{Y.~Salomatin}\thanks{Deceased}   \groupprotvino 
\author{G.~Schnell}  \groupbilbao \groupgent
\author{B.~Seitz}  \groupglasgow
\author{T.-A.~Shibata}  \grouptokyo
\author{M.~Statera}  \groupferrara
\author{E.~Steffens}  \grouperlangen
\author{J.J.M.~Steijger}  \groupnikhef
\author{S.~Taroian}  \groupyerevan
\author{A.~Terkulov}  \groupmoscow
\author{R.~Truty}  \groupillinois
\author{A.~Trzcinski}  \groupwarsaw
\author{M.~Tytgat}  \groupgent
\author{P.B.~van~der~Nat}  \groupnikhef
\author{Y.~Van~Haarlem}  \groupgent
\author{C.~Van~Hulse}  \groupbilbao \groupgent
\author{D.~Veretennikov}  \groupbilbao \groupstpetersburg
\author{V.~Vikhrov}  \groupstpetersburg
\author{I.~Vilardi}  \groupbari
\author{C.~Vogel}  \grouperlangen
\author{S.~Wang}  \groupbeijing
\author{S.~Yaschenko}   \grouperlangen
\author{B.~Zihlmann}  \groupdesy
\author{P.~Zupranski}  \groupwarsaw

\collaboration{The HERMES Collaboration} \noaffiliation


\begin{abstract}
A comprehensive collection of results on longitudinal double-spin asymmetries is presented for charged pions and kaons produced in semi-inclusive deep-inelastic scattering of electrons and positrons on the proton and deuteron, based on the full HERMES data set. The dependence of the asymmetries on hadron transverse momentum and azimuthal angle extends the sensitivity to the flavor structure of the nucleon beyond the distribution functions accessible in the collinear framework.  No strong dependence on those variables is observed. In addition, the hadron charge-difference asymmetry is  presented, which under certain model assumptions provides access to the helicity distributions of valence quarks. 
\end{abstract}

\maketitle

 \section{Introduction}

Measurements of deep-inelastic scattering (DIS) with both beam and target longitudinally polarized
have provided access to the polarization-dependent structure of the nucleon (e.g., Table I of Ref.~\cite{Aidala:2012mv}).
Semi-inclusive DIS, in which an identified final-state hadron is observed in conjunction with the scattered lepton, have provided enhanced sensitivity through the fragmentation process to quark flavor and hence to individual parton distributions \cite{Adeva:1997qz, Ackerstaff:1999ey, Airapetian:2003ct, Airapetian:2004zf, Alekseev:2007vi, Airapetian:2008qf, Alekseev:2009ac,  Alekseev:2010ub, Avakian:2010ae,Jawalkar:2017ube}.
Until recently (e.g., Refs.~\cite{deFlorian:2008mr,Leader:2010rb}),
interpretation of these measurements was largely carried out within a collinear approximation, one for which the effects of transverse components of parton motion are assumed to be negligible.
While yielding substantial knowledge on the longitudinal momentum and polarization structure of the nucleon, such interpretation 
excludes the rich phenomenology of transverse-momentum dependent (TMD) parton distribution and fragmentation functions~\cite{Mulders:1995dh,Bacchetta07}.
In particular, in the limit of small hadron transverse momentum semi-inclusive DIS is  sensitive to intrinsic transverse momentum \cite{Bacchetta:2008xw}.
A detailed theoretical picture has been developed, providing a framework for which semi-inclusive DIS measurements in any configuration of beam and target polarization are related to various combinations of distribution and fragmentation functions~\cite{Mulders:1995dh,Bacchetta07}.

\begin{figure}
	\centering
		\includegraphics[width=0.41 \textwidth]{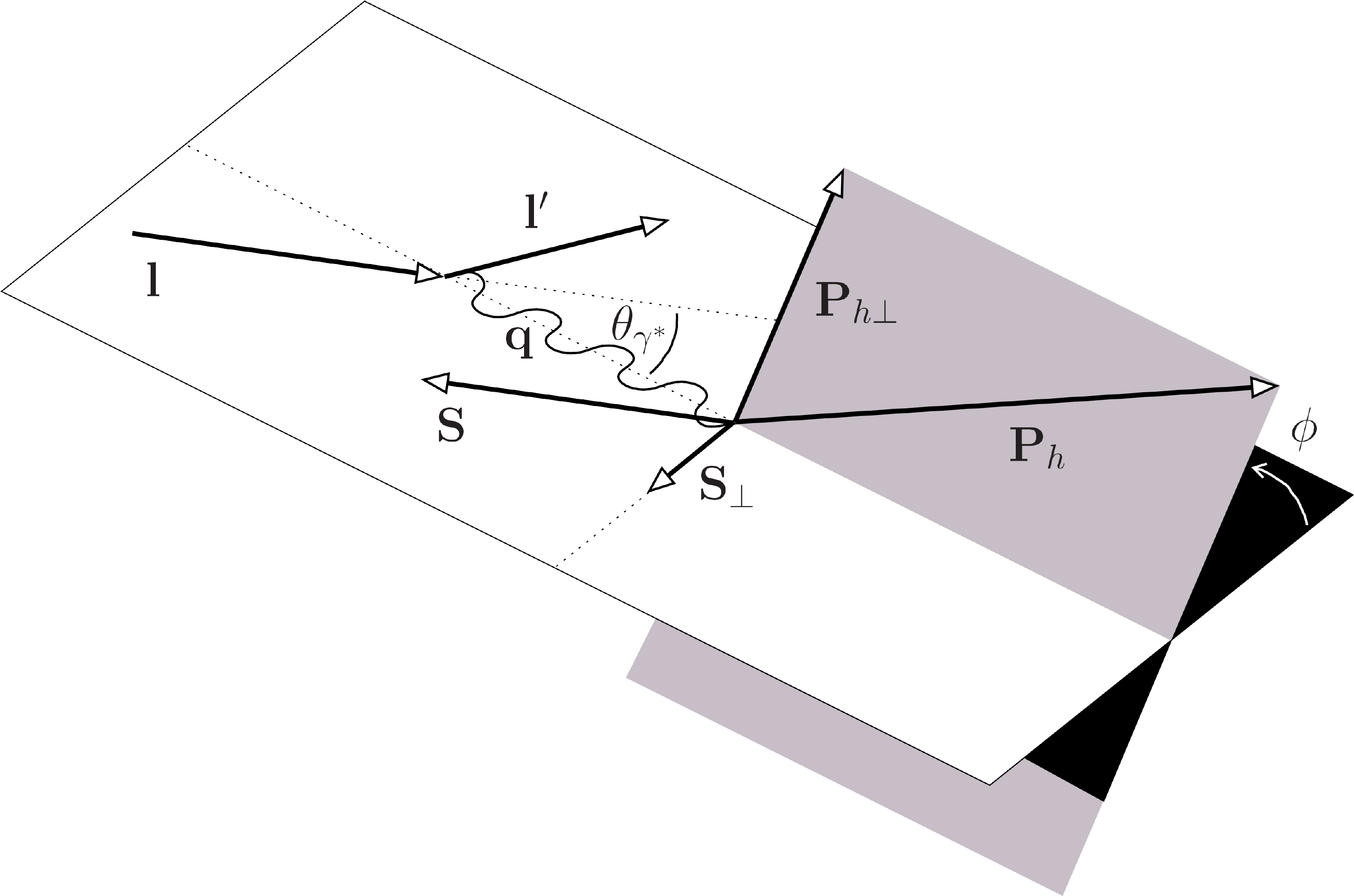}
		\caption{Following the {\em Trento conventions}~\cite{Bacchetta04}, $\phi$ is defined to be the angle between the lepton scattering plane and the plane defined by the virtual-photon momentum \textbf{q} $\equiv $ \textbf{l}$'-$\textbf{l} (the difference of the momenta of the outgoing and incoming lepton) and \textbf{P}$_{h}$, the momentum vector of the observed hadron. \textbf{S} is the spin vector of the nucleon (polarized along the direction of the incoming lepton), while \textbf{S}$_{\perp}$ is its component perpendicular to the virtual-photon direction.}
		\label{fig:paramsDefined}
\end{figure}

If terms that depend on transverse nucleon polarization are neglected, the complete model-independent decomposition of the semi-inclusive DIS cross section in the one-photon--exchange approximation can be expressed in terms of moments of azimuthal modulations  \cite{Bacchetta07},
\begin{align}
\frac{\de\sigma^{h}}{\de x \de y \de z \de P^2_{h\perp} \de{\phi}} = \frac{2\pi\alpha^2}{xyQ^2} \frac{y^2}{2(1-\epsilon)} \bigg( 1 + \frac{\gamma^2}{2x}\bigg) &  \nonumber \\
\bigg\{ F_{UU,T}^{h} + \epsilon F_{UU,L}^{h}  + \lambda \Lambda \sqrt{1-\epsilon^2} F_{LL}^{h}~~~~~~~~~~~~~~~~& \nonumber \\
+ \sqrt{2\epsilon} \, \Big[   \lambda \sqrt{1-\epsilon} \, F^{h,\sin \phi}_{LU}  + \Lambda \sqrt{1+\epsilon} \, F^{h,\sin\phi}_{UL}\Big]  & \sin\phi \nonumber \\ 
+ \sqrt{2\epsilon} \, \Big[ \lambda \Lambda \sqrt{1-\epsilon} \, F^{h, \cos \phi}_{LL} +  \sqrt{1+\epsilon} \, F^{h,\cos\phi}_{UU} \Big]  & \cos\phi \nonumber \\
+ \Lambda\epsilon \, F^{h,\sin2\phi}_{UL}   \sin2\phi + \epsilon  \, F^{h,\cos2\phi}_{UU}  \cos2\phi~\bigg\}&.
\label{eqn:big-cross}
\end{align}
The variables $Q^2$, $y$, and $x$ are the negative squared four-momentum of the virtual photon, the fraction of beam energy carried by the virtual photon in the target rest frame, and the Bjorken scaling variable, respectively. Here, \(x=Q^2/(2M\nu)\) with $M$ the mass of the proton and $\nu$ the energy of the virtual photon in the target rest frame. 
These variables are determined from the momentum and angle of the scattered lepton.
The angle $\phi$ is the azimuthal angle of the hadron momentum vector \textbf{P}$_{h}$ about the virtual-photon direction with respect to the lepton-scattering plane as depicted in Fig.~\ref{fig:paramsDefined} and defined, e.g., in Ref.~\cite{Bacchetta04}. 
The  $F^{h,\text{mod}}_{XY,Z}$ of Eq.~\eqref{eqn:big-cross} represent  structure functions whose subscripts denote the polarization of the beam, of the target (with respect to the virtual-photon direction), and---if applicable---of the virtual photon. The superscript indicates the dependence on the hadron type and the azimuthal modulation parametrized. Each of these structure functions is a function of $x$, $Q^2$, $z$, and $P_{h\perp}$, where $z$ is the fraction of the virtual-photon energy carried by the observed final-state hadron (in the target rest frame), while  $P_{h\perp}$ is the magnitude of the hadron momentum component transverse to the virtual-photon direction.   
The helicity of the nucleon in the center-of-mass system of the virtual photon and the nucleon is denoted as $\Lambda$, while
$\lambda$ represents the helicity of the beam lepton.
Furthermore, the ``photon polarization parameter'' $\epsilon = \frac{1-y-\frac{1}{4}~\gamma^{2}~y^2}{1-y+\frac{1}{4}~y^{2}~(\gamma^2+2)}$ is the ratio of longitudinal-to-transverse photon flux, where $\gamma= Q/\nu$, 
and $\alpha$ is the fine-structure constant.

In order to probe the polarization-dependent structure of the nucleon with minimal experimental systematic uncertainties, spin asymmetries are typically measured instead of cross sections.  
Ideally, cross sections are compared in all combinations of 100\% polarized beams (with respect to beam direction) and targets (with respect to virtual-photon direction) to form~\cite{Diehl:2005} 
\begin{equation}
  A_{LL}^{h}\equiv 
	\frac{\sigma^{h}_{+-}-\sigma^{h}_{++}+\sigma^{h}_{-+}-\sigma^{h}_{--}}{\sigma^{h}_{+-}+\sigma^{h}_{++}+\sigma^{h}_{-+}+\sigma^{h}_{--}}.
\label{eqn:ALL}
\end{equation}
Here, $\sigma^{h}_{\lambda\Lambda}$ 
denotes the cross section in a given configuration of equal and opposite beam and target helicities. 
In a typical experimental situation of incomplete polarizations of beam and target, the degrees of polarization of the beam
and target must be divided out. 

The $P_{h\perp}$ dependence of semi-inclusive asymmetries is sensitive
to the transverse-momentum contributions from both 
the partonic structure of the nucleon and the fragmentation process 
through which final-state hadrons are produced.  Transverse-momentum distributions have 
in recent years become topics of great interest. 
The focus has been primarily on their relationship to transverse-spin asymmetries.
However, even for unpolarized or longindually polarized beam/target, the $P_{h\perp}$ dependence has been shown to be sensitive to various sources of transverse momentum in the nucleon~\cite{Anselmino:2006yc,Avakian:2007xa,Musch:2010ka}.

In the limit of small hadron transverse momentum ($P_{h\perp}\ll zQ$), the various contributions to $A^{h}_{LL}(x,Q^2,z,P_{h\perp},\phi)$ can be expressed in terms of convolutions of TMD distribution with fragmentation functions.  
The azimuthally uniform $A^{h}_{LL}(x,Q^2,z,P_{h\perp})$ enters with a single leading-twist contribution:
\begin{equation}
F^{h}_{LL} \propto \sum_q e_q^2 \bigg[~g_{1L}^q (x,p_T^2) \otimes_{\mathcal{W}_1} D_1^{q\to h}(z,k_T^2)~\bigg].
\end{equation}
Here,  ``$\otimes_{\mathcal{W}_1}$'' represents a convolution of the distribution and fragmentation functions over the intrinsic transverse momentum $p_T$ of the parton $q$ (with fractional charge \(e_q\)) and the transverse-momentum contribution $k_T$ from the fragmentation process  with a kinematic ``weight'' $\mathcal{W}_1$.   
The function $\mathcal{W}_1$ [and $\mathcal{W}_2$ from Eq.~\eqref{eq:subleading-TMD}] is given explicitly, e.g., in Ref.~\cite{Bacchetta07}. 
In the collinear limit, $F^{h}_{LL}$ reduces to the well-known product of the collinear helicity distribution $g^{q}_{1}(x)$ and the collinear fragmentation function $D_{1}^{q\to h}(z)$.\footnote{The additional but suppressed contribution related to the $g_{2}$ structure function is neglected here due to the smallness of $g_{2}$ (e.g., Ref.~\cite{Airapetian:2011wu}).}

While there are no possible azimuthal moments at leading twist, cosine modulations are potentially present at twist-three level, i.e., suppressed by a single power of $M/Q$.  Taking the Wandzura--Wilczek approximation (neglecting interaction-dependent terms, which depend on quark-gluon-quark correlators, and neglecting terms linear in quark masses)~\cite{Wandzura:1977qf}, the following expression remains:
\begin{align}
F_{LL}^{h,\cos\phi} \propto \frac{M}{Q}\sum_q e_q^2 \bigg[~g_{1 L}^{q} (x,p_T^2) & \otimes_{\mathcal{W}_2} D_1^{q\to h} (z,k_T^2)~\bigg]
\label{eq:subleading-TMD}.
\end{align}
This combination of distribution and fragmentation functions was studied, e.g., in Ref.~\cite{Oganessyan:2002pc}, and is sometimes referred to as the ``polarized Cahn effect'', which combines transverse momentum of longitudinally polarized partons inside the target nucleon with transverse momentum produced in the fragmentation process.

The unpolarized denominator of Eq.~\eqref{eqn:ALL} has been extensively studied. Kinematic dependences of its azimuthal modulations, which include contributions arising from, e.g., the {\em Boer--Mulders}~\cite{Boer:1997nt} and {\em Cahn}~\cite{Ravndal73, Kingsley74, NCahn1978269} effects, have been explored thoroughly in Refs.~\cite{Airapetian:2012yg,Adolph:2014pwc,Yan:2016ods}.

In general, the use of Eq.~\eqref{eqn:ALL} to extract information on the nucleon spin structure in terms of parton distributions requires knowledge of the hadronization process.
The advantage of such information is a more detailed sensitivity to the various quark flavors than that of purely inclusive DIS.

The hadron charge-difference double-spin asymmetry provides additional spin-structure information and is not trivially constructible from the simple semi-inclusive asymmetries. Under certain symmetry assumptions for fragmentation functions (cf.~Sec.~\ref{sec:charge-difference}) charge-difference asymmetries provide a direct extraction of valence-quark polarizations~\cite{Frankfurt:1989wq}.

It is the primary goal of this article to present the kinematic dependences of 
hadron-tagged longitudinal double-spin asymmetries as completely as possible with the available data.
In comparison to the analysis of the HERMES unpolarized data presented in Ref.~\cite{Airapetian:2012yg}, the size of the data presented here did not allow for a complete five-dimensional kinematic unfolding of the data. Decisions were made about the best possible kinematic projections of these data, within the constraints of the theoretical framework described above and with the goal of providing the maximum possible access to physics of interest.

\emph{For the purpose of discussion and comparison} some assumptions will be made in the analysis, but the lepton-nucleon asymmetry $A^{h}_{\parallel}$ [cf.~Eq.~\eqref{eq:a-parallel-born}] should be taken to be the primary model-independent observable and is provided in the data tables in all cases. This asymmetry differs from $A^{h}_{LL}$ only in the direction in which the nucleon polarization is measured, either with respect to the beam for the former, or with respect to the virtual photon for the latter. Because of this, $A^{h}_{\parallel}$ contains a relatively small, but non-vanishing component of $A^{h}_{LT}$ \cite{Diehl:2005}.\footnote{The polarization component transverse to the virtual-photon direction (see Fig.~\ref{fig:paramsDefined}) is proportional to $\sin\theta_{\gamma^{*}}$, where $\theta_{\gamma^{*}}$ is the angle between the incoming lepton momentum and the virtual-photon direction. This transverse component is 10\%-15\% of the target polarization in typical HERMES kinematics, but can reach 20\% for the largest $x$ values covered in this experiment.} 
More significantly, in order to relate $A^{h}_{\parallel}$ to the virtual-photon--nucleon asymmetry $A^{h}_1$ [cf.~Eq.~\eqref{eq:A1}], a parameterization of the longitudinal-to-transverse photoabsorption cross-section ratio $R=\sigma_L/\sigma_T$ must be assumed. To date, this quantity has only been measured in inclusive DIS.  However, in semi-inclusive DIS this ratio might depend strongly on the hadron kinematics, in particular on $P_{h\perp}$.

\section{Measurement and analysis}

\subsection{HERMES experiment and analysis formalism}

	 The data were collected using the HERMES spectrometer 
	\cite{Ackerstaff:1998av} at the HERA storage ring during the 1996--2000 
	running period. A longitudinally polarized lepton (electron or positron) 
	beam with a momentum of 27.6~GeV was scattered off a longitudinally polarized atomic hydrogen or 
	deuterium gas target. 
	The sign of the target polarization was randomly chosen each 60~s for hydrogen and 90~s for deuterium, 
	providing yields in both spin states while controlling systematic uncertainties. 
	The experimental configurations by year are summarized in Table~\ref{tab:configByYear}. 
	Typical values for the beam (target) polarization are around 53\% (84\%).

	\begin{table}
     \begin{ruledtabular}
	\caption{
		Experimental configurations by year of longitudinally polarized beam and 
		target data taking. The varieties of hadrons identified and the 
		hadron-momentum range 
		are determined by the particle-identification systems 
		available at the time. A threshold Cherenkov counter was used during 
		the hydrogen data-taking period and a ring-imaging Cherenkov 
		detector was used throughout the deuterium 
		running period. 
		}
	\label{tab:configByYear}
	\begin{tabular}{ccccc}
	    ~   & ~Beam~ & ~Target~ 	& ~Hadron~		& Hadron Momentum \\
	Year &  Type 	&   Gas   	&  Type 		& $P_h$\\
	\hline
	1996 & e$^+$	& H 	&  $\pi^{\pm}$			& 4--13.8 GeV \\
	1997 & e$^+$	& H 	&  $\pi^{\pm}$			& 4--13.8 GeV \\
	1998 & e$^-$	& D 	&  $\pi^{\pm}, K^{\pm}$	& 2--15    GeV \\
	1999 & e$^+$	& D 	&  $\pi^{\pm}, K^{\pm}$	& 2--15    GeV \\
	2000 & e$^+$	& D 	&  $\pi^{\pm}, K^{\pm}$	& 2--15    GeV \\
	\end{tabular}
     \end{ruledtabular}
	\end{table}
	
	The asymmetries are computed using basically the same data set and procedure presented 
	in prior HERMES publications on longitudinal double-spin asymmetries 
	\cite{Ackerstaff:1999ey, Airapetian:2003ct, Airapetian:2004zf, Airapetian:2007mh};
	differences from previous analyses are discussed below.  
	The lepton-nucleon asymmetry is 
	\begin{equation}
	A^{h}_{\parallel}  \equiv \frac{C^{h}_{\phi}}{f_D}\left[\frac{L_{\rightrightarrows}N^{h}_{\rightleftarrows} - 
		L_{\rightleftarrows}N^{h}_{\rightrightarrows}}{L_{P,\rightrightarrows} 
		N^{h}_{\rightleftarrows} + 
		L_{P,\rightleftarrows}N^{h}_{\rightrightarrows}}\right]_\text{B}  .
	\label{eq:a-parallel-born}
	\end{equation}
	Here, $N^{h}_{\rightrightarrows(\rightleftarrows)}$ represents the hadron yield
	 containing events that meet the kinematic requirements
	summarized in Table~\ref{tab:cuts}, and $L_{\rightrightarrows(\rightleftarrows)}$ and
	$L_{P,\rightrightarrows(\rightleftarrows)}$ represent the luminosity 
	and polarization-weighted luminosity in the parallel (antiparallel) 
	experimental beam/target helicity configuration.\footnote{Note that if experimental polarizations are not alternated so that the average polarization of both beam and target samples are zero, terms in Eq.~\eqref{eqn:big-cross} with a single ``$U$'' in the subscript do not vanish, a priori, from both the numerator and denominator of the ratio. In contrast, Eq.~\eqref{eqn:ALL}, i.e., the combination of all four target- and beam-helicity states, leaves only the sum of terms from Eq.~\eqref{eqn:big-cross} with the ``$LL$'' subscript divided by the sum of terms with the ``$UU$'' subscript.}
	The square brackets, $\left[~\right]_\text{B}$, 
	indicate that the enclosed quantity is 
	corrected to Born level, i.e., unfolded for radiative and detector 
	smearing, using Born and smeared Monte Carlo simulations according to 
	the essentially model-independent procedure described in Ref.~\cite{Airapetian:2004zf}.
	The unfolding is carried out in the same dimension used to present the data 
	(see also Section~\ref{sec:results} and Table~\ref{tab:binning}). 
	The factor $f_D$ represents the dilution of the polarization of the nucleon with respect to
	that of the nucleus and is explained in Section~\ref{sec:fD}.  Finally, $C^{h}_{\phi}$ is a correction
	that compensates for any distortion caused by the convolution of the azimuthal
	moments of the polarization-independent cross section with the non-uniform detector
	acceptance, which is described in more detail in Section~\ref{sec:azimcorr}.

	\begin{table}    
		\caption{
			Inclusive and semi-inclusive kinematic requirements (value in parentheses is the limit for the {\em extended} range discussed in Section~\ref{sec:min-z}). Here, Feynman-$x$ (\(x_{F} \)) is defined as the ratio of the hadron's longitudinal momentum component in the virtual-photon--nucleon center-of-mass system to its maximal possible value.  
			} 
		\label{tab:cuts}
		\centering
		\begin{ruledtabular}
        \begin{tabular}{c}
		Kinematic Requirements\\
		\hline
		$Q^2  > 1.0$ GeV$^{2}$ \\
		$W^2 > 10$ GeV$^2$\\
		$y < 0.85$\\  
		(0.1) $0.2<z<0.8$ \\
		$x_{F} > 0.1$ \\
        \end{tabular}
		\end{ruledtabular}
	\end{table}

	The virtual-photon--nucleon asymmetry $A_1^{h}$ is defined as
	\begin{align}
		A_1^h&\equiv 
		\frac{\sigma^{h}_{1/2}-\sigma^{h}_{3/2}}{\sigma_{1/2}^{h}+\sigma_{ 
		3/2}^{h}}, 
		\label{eq:A1} 
	\end{align} 
	where $\sigma^{h}_{1/2}$ ($\sigma^{h}_{3/2}$) is the photoabsorption cross section for photons for which the spin is antiparallel (parallel) to the target-nucleon spin.
	$A_1^{h}$ is computed from $A^{h}_{\parallel}$ as 
	\begin{align}
		A_1^h &= \frac{1}{D(1+\eta\gamma)}A_{\parallel}^{h},
        \label{eq:A1exp} 
	\end{align} 
	where the contributions from the spin structure function $g_{2}$ and, in case of a deuterium target, from the tensor structure function $b_{1}$ are negligible~\cite{Airapetian:2005cb}. Furthermore,
	\begin{equation}
	\eta = 
	\frac{\epsilon\gamma y}{1-\left(1-y\right)\epsilon }\label{eq:eta}
	\end{equation}
	 is a kinematic factor, and
	\begin{equation}
		D = \frac{1-(1-y)\epsilon}{1 + \epsilon R}
	\end{equation}
	accounts for the limited degree of polarization transfer at the 
	electron--virtual-photon vertex, including the ratio $R$ of longitudinal-to-transverse cross sections.
	In this analysis, 
	$R$ was taken from the R1999 parameterization 
	\cite{collaboration-1999-452} for all calculations of $A^{h}_1$, which---strictly speaking---is valid only for inclusive DIS measurements as pointed out above.

\subsection{Differences from prior analyses}\label{sec:analysis-differences}

	Although the analysis has much in common with those in prior 
	HERMES publications, several changes are made, which increase 
	statistical precision and reduce the systematic uncertainties.

	\subsubsection{Nucleon-polarization correction}\label{sec:fD}
	
	The factor $f_D$ in Eq.~\eqref{eq:a-parallel-born} is the ratio of the 
	polarization of the target nucleon to that of the host nucleus. 
	This value is unity for protons,
	and 0.926 for deuterons due to the D-state admixture in the deuteron wave function~\cite{physRevC.60.035201}. 
	The application of this correction directly to the 
	asymmetries differs from the analysis of Ref.~\cite{Airapetian:2004zf}. 
	In this prior publication, the nucleon 
	polarization correction was applied in a calculation of quark 
	polarizations but not to the asymmetries themselves.

	\subsubsection{Minimum-$z$ requirement}\label{sec:min-z}
	As in prior analyses, a constraint on the hadron energy-fraction of $z>0.2$ is applied.
	Only for the two-dimensional binning performed in $x$ and $z$, an additional low-$z$ bin (0.1$< z <$ 0.2) is 
	added, which provides access to the kinematic behavior of the asymmetry outside the region that is typically 
	used to separate current and target-remnant regions. This bin is omitted, 
	however, from polynomial fits of Table~\ref{tab:xz-chisq} as the fit is intended as a check in a $z$ range 
	commonly used in global analyses.

	\subsubsection{Minimum-$P_h$ requirement} 
	
	The hadron momentum range accepted is determined by the capabilities of 
	the hadron identification apparatus. For the hydrogen sample, a threshold 
	Cherenkov counter requires a minimum hadron momentum of 4~GeV in 
	order to distinguish charged pions from heavier hadrons. For the 
	deuterium sample the installation of a dual-radiator 
	ring-imaging Cherenkov detector (RICH)~\cite{Akopov:2000qi}
	enabled identification of hadrons with momentum
	larger than 2~GeV. For historical reasons, prior asymmetry analyses 
	required the minimum hadron momentum to be the same for the two targets. 
	This restriction has been relaxed on the deuterium sample as it
	unnecessarily removes low-momentum hadrons.

	\subsubsection{Event-level RICH unfolding}
	
	In comparison to prior analyses, the RICH hadron identification algorithm was 
	improved to reconstruct better multi-hadron events~\cite{Airapetian:2012yg}.  For each 
	event, hit patterns are produced for each possible combination of hadron 
	hypotheses so that the effect of all tracks is taken into account simultaneously.  
	Previously each hadron track was identified individually, which increased the 
	probability of misidentification for cases where Cherenkov rings overlapped.

	\subsubsection{Multidimensional unfolding}
	
	Event migration due to radiative and detector smearing is corrected for in an
	unfolding procedure as in the previous HERMES analyses. The exploration
	of multidimensional dependences in this analysis required unfolding
	not only in $x$ but also in the other variables under study (s.~below).
	However, unlike the case of unfolding of polarization-independent 
	hadron yields in the measurement of hadron multiplicities~\cite{Airapetian:2012ki},
	which are strongly dependent on the hadron kinematics and which thus require 
	also unfolding in those variables, the dependence of longitudinal double-spin asymmetries 
	on hadron kinematics in this analysis is weak and the unfolding procedure
	is found to be robust against possible model dependence when performed
	in only those dimensions presented here, e.g., when the polarization-dependent 
	yields are integrated over $z$ or $P_{h\perp}$.

	\subsubsection{Azimuthal-acceptance correction}\label{sec:azimcorr}
	
	The factor $C^{h}_{\phi}$ in Eq.~\eqref{eq:a-parallel-born} 
    is a correction applied to the semi-inclusive 
	asymmetries that compensates for the influence of the spectrometer 
	acceptance in the implicit integration over kinematic variables in the semi-inclusive
	yields. It is primarily the integral over $\phi$, which combines a non-uniform 
	detector acceptance with azimuthal modulations in the polarization-independent yield, 
	produced, for example, by the Cahn effect~\cite{NCahn1978269}, which 
	distorts the semi-inclusive asymmetries. In practice, 
	the \emph{actual} asymmetry that is measured, $\tilde{A}_{\parallel}^h$, involves a 
	convolution with an acceptance function $\xi(\phi)$, such that 
	\begin{align}
		\tilde{A}_{\parallel}^h(x&, Q^2, z, P_{h\perp}) \nonumber \\
		 &= \frac{\int 
		\de \phi~\sigma_{\parallel}^h\left(x,Q^2,z,P_{h\perp},\phi\right)~\xi(\phi) 
		}{\int 
		\de \phi~\sigma_{UU}^h\left(x,Q^2,z,P_{h\perp},\phi\right)~\xi(\phi)}. 
		\label{eq:acc} 
	\end{align}
	
	In order to correct for this effect in the denominator of the asymmetry,
	a recent parameterization of the azimuthal moments of HERMES unpolarized data~\cite{Airapetian:2012yg} was used.  
	This  parameterization was produced by unfolding
	unpolarized semi-inclusive yields in all five kinematic degrees of freedom simultaneously.
	The unfolding was conducted simultaneously in 10800 (5$x~\times$ 5$y~\times$ 6$z~\times$ 6$
	P_{h\perp}~\times$ 12$\phi$) bins, correcting the 
	measured yields for acceptance and smearing effects.

	The unpolarized correction factor,
	\begin{equation}
		C_\phi^h =  \frac{\mathcal{A}_{\parallel}^h}{\tilde{\mathcal{A}}_{\parallel}^h},
	\end{equation}
	is formed by taking the ratio of two Monte Carlo simulated asymmetries computed in acceptance: $\mathcal{A}_{\parallel}^h$, 
	which is generated without azimuthal modulations (e.g., Cahn and Boer--Mulders effects), and $\tilde{\mathcal{A}}_{\parallel}^h$, 
	which is weighted event-by-event by the parameterized azimuthal modulation
	of the polarization-independent cross section~\cite{Airapetian:2012yg},
	to reproduce the effect of the non-uniform azimuthal acceptance. By applying this 
	ratio, the unpolarized denominator of the measured asymmetry is corrected for 
	azimuthal acceptance effects.
	This correction is typically less than a few percent, reaching (and occasionally exceeding) 10\% only in the kinematic region of large $x$.
	
	The polarization-dependent numerator of Eq.~\eqref{eq:acc} is also subject to 
	possible azimuthal modulations, which can enter the cross section at 
	subleading twist~\cite{Bacchetta:2008xw}. Sizable subleading-twist effects have 
	in fact been observed in unpolarized-beam, longitudinally polarized target 
	asymmetries \cite{Airapetian:2005jc}, which underscores the need to proceed 
	with some caution.  As the unbinned yields are limited in the  
	dataset for a complete five-parameter kinematic unfolding, a full parameterization of 
	polarization-dependent modulations is not possible, preventing a 
	correction similar to that described for the unpolarized azimuthal acceptance. In 
	order to address this, but also to access these additional degrees of freedom in the 
	polarization-dependent cross section, $A^{h}_{\parallel}$ was unfolded simultaneously in $x$ and $\phi$.
	The \( \cos\phi \) and \( \cos2\phi \) moments of $A^{h}_{\parallel}$, which will be presented
	in Section~\ref{sec:azimresults}, were found to be consistent with zero.

	\subsubsection{Analytic fits}
	
	The two-dimensionally ($x$-$P_{h\perp}$ and $x$-$z$) binned virtual-photon--nucleon asymmetries $A^{h}_{1}$  are simultaneously fit
	with polynomial functions in both dimensions.
	This has two significant benefits. First, as kinematic variables are correlated
	to some degree, fitting provides a means of separating the underlying kinematic
	dependences of the asymmetries from kinematic correlations.  Different hypotheses
	for kinematic dependence can easily be compared on the basis of their goodness-of-fit.
	As an example, a weak though non-vanishing dependence on $P_{h\perp}$ of $A^{h}_{1}$ has been suggested by lattice-QCD calculations~\cite{Musch:2010ka},
	which would have to be disentangled from the much stronger dependence on $x$.
	Second, such fits present a more intuitive picture of the
	statistical significance of data for which there are large covariances between bins.
	As is the case when a model-independent radiative and detector-smearing unfolding procedure
	is applied, some inflation of the on-diagonal error matrix elements occurs.  While this causes 
	the uncertainties to appear larger, the effect is compensated for by the statistical correlations between
	bins~\cite{Airapetian:2004zf,Airapetian:2007mh}.
	By presenting a fit in addition to the data points with their  single-bin
	uncertainties, the statistical power of the data to constrain models is also conveyed.

\section{Results}\label{sec:results}	

\subsection{One- and two-dimensional projections of $A_{\parallel}^h$}

	The leading contribution to the longitudinal double-spin asymmetry~\eqref{eqn:ALL} is the azimuthally uniform $A_{\parallel}^{h}(x,z,P_{h\perp})$. Traditionally, its collinear version, i.e., integrated over $P_{h\perp}$, has been presented as a function of $x$ only as the dependence on $z$ through the spin-independent fragmentation functions in the numerator and denominator largely cancels. Nevertheless, further information on the underlying interplay of parton distribution and fragmentation functions can be obtained by analyzing the multi-dimensional dependences. They are, in addition, less prone to potential detector effects that arise from integration of the numerator and denominator of Eq.~\eqref{eq:acc} separately over the larger region of phase space on which the detector acceptance and physics observable might depend. 
	
	In this analysis the polarization-dependent experimental hadron yields were corrected for radiative and detector smearing by an unfolding procedure as described in Ref.~\cite{Airapetian:2004zf}. As pointed out already, in contrast to the earlier analysis, unfolding was performed not only in one dimension, $x$, but also in two or three dimensions. Due to limited yields, the binning in the kinematic variables differs for each of the projections chosen, with the highest resolution in $x$ for the one- and three-dimensional presentations. 

\begin{table}[t]
	\caption{
			Bin boundaries used for the various presentations of $A_{||}^{h}$.
			}
	\label{tab:binning}
	\centering
\footnotesize	
	\begin{tabular}{lc}
		\hline 
		\hline
		\multicolumn{2}{c}{one-dimensional binning in $x$} \\
		\hline
		\multicolumn{2}{c}{0.023 -- 0.04 -- 0.055 -- 0.75 -- 0.1 -- 0.14 -- 0.2 -- 0.3 -- 0.4 -- 0.6} \\[0.4cm]
		\hline 
		\hline
		\multicolumn{2}{c}{two-dimensional binning in $x$ and $z$} \\
		\hline
		$x$: & 0.023 -- 0.055 -- 0.1 -- 0.6 \\		
		$z$: & 0.1 -- 0.2 -- 0.3 -- 0.4 -- 0.5 -- 0.6 -- 0.7 -- 0.8\\[0.4cm]
		\hline 
		\hline
		\multicolumn{2}{c}{two-dimensional binning in $x$ and $P_{h\perp}$} \\
		\hline
		$x$: & 0.023 -- 0.055 -- 0.1 -- 0.6 \\		
		\multicolumn{2}{l}{$P_{h\perp}$ [GeV]: \hspace*{.55cm}  0 -- 0.15 -- 0.3 -- 0.4 -- 0.5 -- 0.6 -- 2.0}\\[0.4cm]
		\hline 
		\hline
		\multicolumn{2}{c}{three-dimensional binning in $x$, $P_{h\perp}$, and $z$}\\
		\hline
		$x$:~  & 0.023 -- 0.04 -- 0.055 -- 0.75 -- 0.1 -- 0.14 -- 0.2 -- 0.3 -- 0.4 -- 0.6 \\		
		\multicolumn{2}{l}{$P_{h\perp}$ [GeV]: \hspace*{1.55cm}  0 -- 0.3  -- 0.5  -- 2.0}\\
		$z$:  & 0.2 -- 0.35 -- 0.5 -- 0.8 \\
		\hline
	\end{tabular}
\end{table}

	\begin{figure*}
		\centering
		\includegraphics[width=0.68 \textwidth]{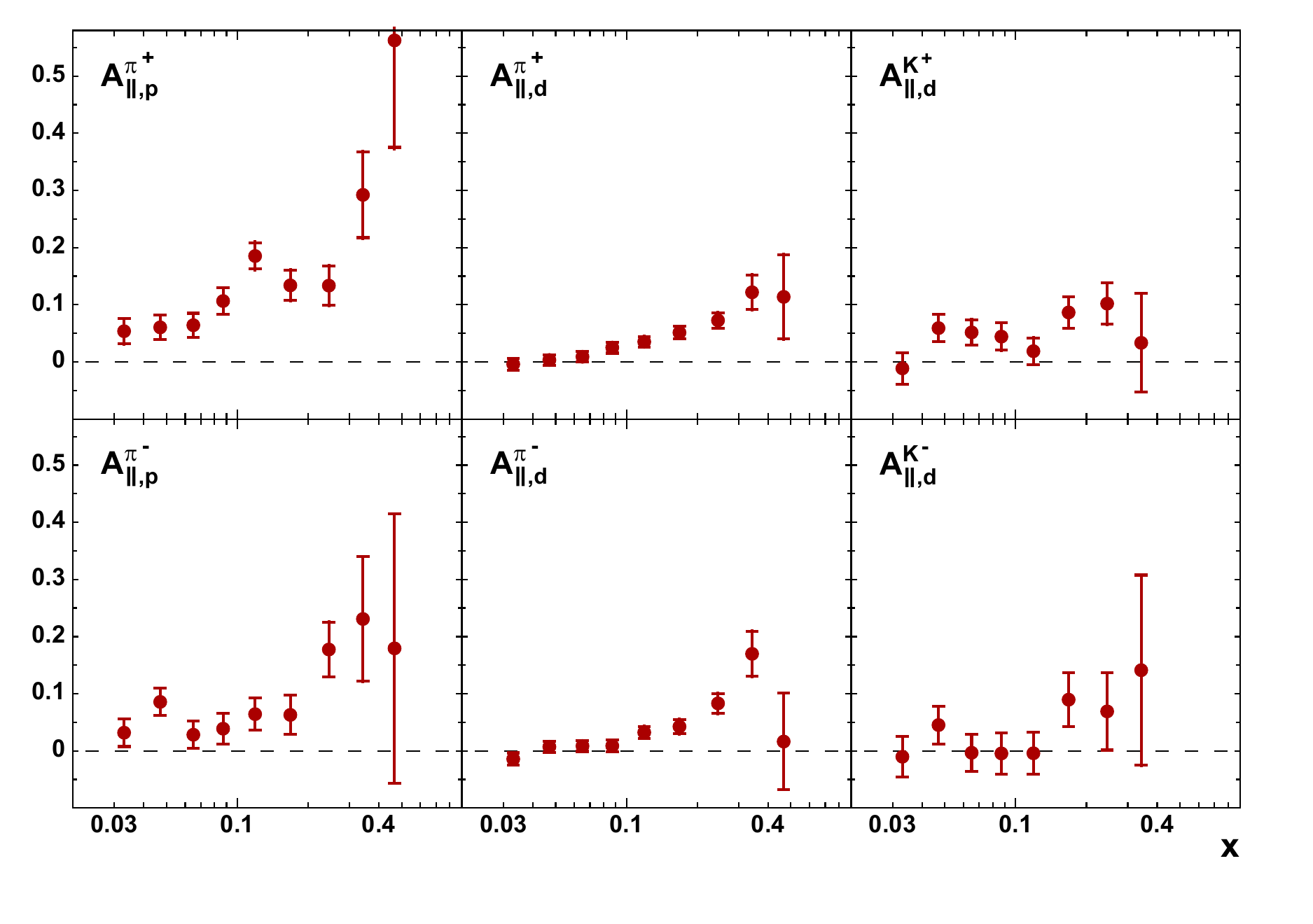}
		\caption{
			The longitudinal double-spin asymmetries $A_{\parallel, N}^h$ as a function of $x$
			with $N=p,d$ denoting the target nucleus and $h=\pi^{\pm},K^{\pm}$ the final-state hadron detected.
			The inner error bars represent statistical uncertainties while the outer ones statistical and systematic 
			uncertainties added in quadrature (hardly visible in this figure). 
			}
		\label{fig:lowQ2d}
	\end{figure*}

The dominating systematic uncertainty stems from the knowledge of both beam and target polarization, 
and amounts to an average relative uncertainty of 
6.6\% for the hydrogen and 
5.7\% for the deuterium data.
Contributions to the systematic uncertainties from the RICH as well as acceptance and smearing unfolding were found to be substantially smaller than those.
Contributions from the azimuthal-acceptance correction amount up to about 3\% at large $x$ while becoming negligible at small $x$. 
The total systematic uncertainty, quoted in the data tables, is the quadratic sum of all contributions. 
In the figures they are added in quadrature to the statistical uncertainty.

	In order to produce asymmetries in a fine binning in $x$, 
	yields were binned in two dimensions: $x$ and two ranges in $Q^2$.
	The low-$Q^2$ bin was added, spanning 0.5 to 1~$\text{GeV}^2$,
	to allow for a better control of migration of events in the unfolding procedure. 
	Likewise, the $x$ region of interest was subdivided into nine bins 
	(see Table~\ref{tab:binning}), 
	with again additional ``padding'' bins at low $x$.
	This quasi two-dimensional binning made it possible to perform 
	kinematic unfolding (as described above) in $x$ and $Q^2$ 
	simultaneously, which compensated for events that migrated from one joint 
	$x$-$Q^2$ bin to another due to QED radiative corrections or detector smearing.

The resulting $x$ dependence of the asymmetries is presented for hydrogen and deuterium targets in Fig.~\ref{fig:lowQ2d}.
The asymmetries extracted were found to be essentially identical to those in prior HERMES analyses~\cite{Airapetian:2004zf}.


\begin{figure*}
	\centering
	\includegraphics[width=0.67 \textwidth]{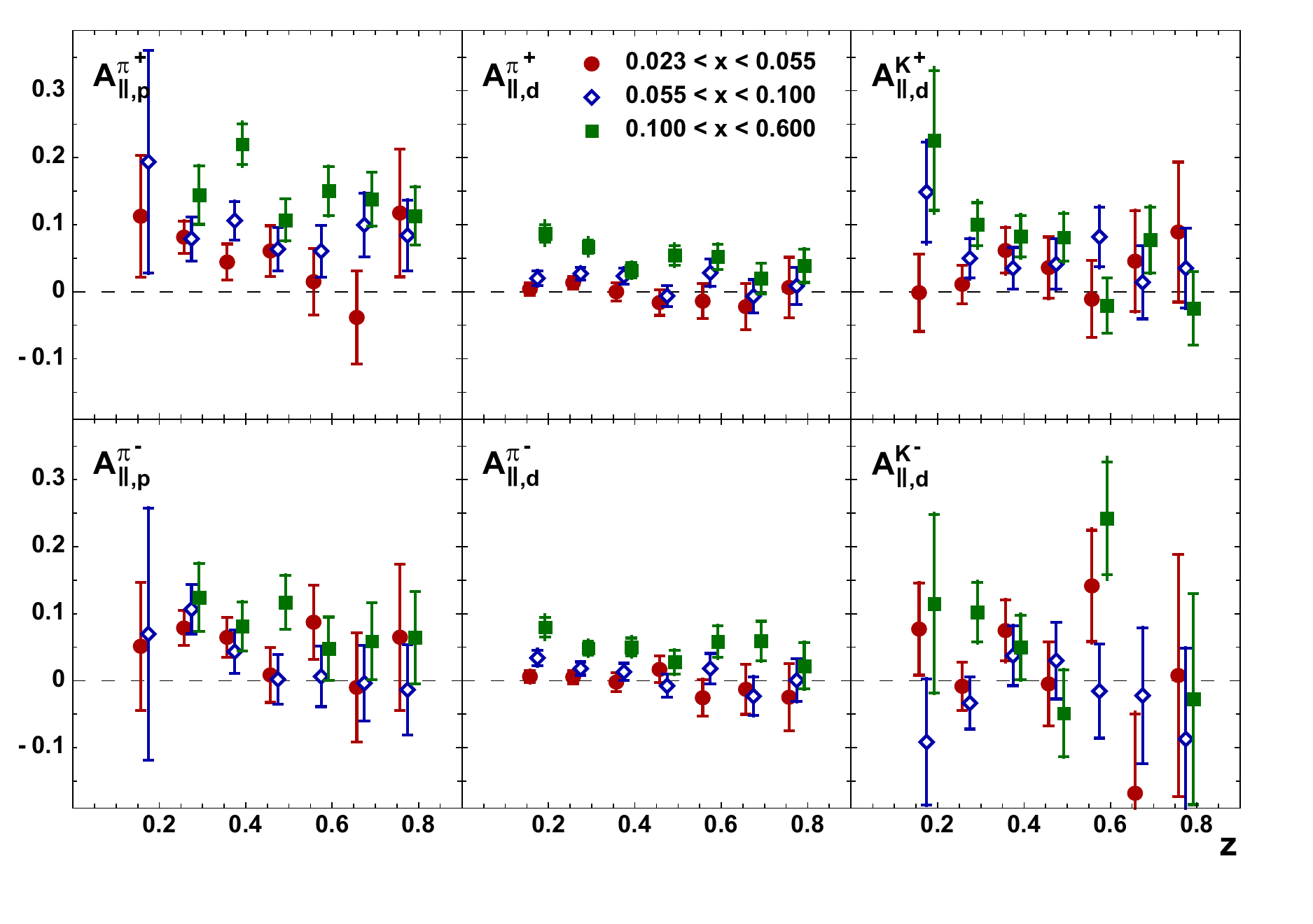}
	\caption{
			The longitudinal double-spin asymmetries $A_{\parallel, N}^h$ as a function of $z$ in three different $x$ ranges as labeled, 
			with $N=p,d$ denoting the target nucleus and $h=\pi^{\pm},K^{\pm}$ the final-state hadron detected.
			Data points for the first $x$ slice are plotted at their average kinematics, while the ones for 
			the other two $x$ slices are slightly shifted horizontally for better legibility.
			The inner error bars represent statistical uncertainties while the outer ones 
			statistical and systematic uncertainties added in quadrature.
			}
	\label{fig:2DXZ}
\end{figure*}

\begin{table*}
     \begin{ruledtabular}
	\caption{
			The $\chi^2$ values for polynomial fits to the $A^{h}_{1,N}(x,z)$ 
			data points for each combination of target (\(N=p,d\)) and final-state hadron \(h\), 
			and number of degrees of freedom ($NDF$) as indicated. The $0.1<z<0.2$  bin 
			has been excluded from fits in order to test for $z$-dependence in the region 
			commonly used in global analyses.  
            The $C_i^h$ are the polynomial terms of the fit functions.
			Except where clearly over-parameterized,
			the fit function linear in $z$ yields little improvement over the fit  constant in that variable 
			suggesting little or no $z$ dependence of the asymmetry. 
			}
	\label{tab:xz-chisq}
	\begin{tabular}{lcccccc}
		~& \( A^{\pi^{+}}_{1,p} \) & \( A^{\pi^{-}}_{1,p} \) & \( A^{\pi^{+}}_{1,d} \) & \( A^{\pi^{-}}_{1,d} \) & \( A^{K^{+}}_{1,d} \) & \( A^{K^{-}}_{1,d} \)   \\[.5mm]
		\hline 
		$\chi^{2~(NDF=16)}_{~C_1^h+C_2^hx}$ & 12.6 & 10.0 & 13.4 & 9.1 & 10.7 & 26.0 \\ [3mm]
		$\chi^{2~(NDF=15)}_{~C_1^h+C_2^hx+C_3^hz}$ & 12.2 & 6.3 & 7.2 & 7.2 & 10.1 & 24.8 \\ [3mm]
		$\chi^{2~(NDF=12)}_{~C_1^h+C_2^hx+C_3^hz+C_{4}^{h} x^{2} + C_{5}^{h} z^{2} +C_{6}^{h} xz}$ & 10.3&4.5&5.5&4.8&5.8&16.1\\[.5mm]
	\end{tabular}
     \end{ruledtabular}
\end{table*}

The $z$ dependence of fragmentation functions is in principle quark-flavor dependent. This can result in an additional dependence of $A_{\parallel}^{h}$ on $z$. Nevertheless, the $z$-dependence of longitudinal double-spin asymmetries is a largely unexplored degree of freedom. This is addressed in a two-dimensional analysis, in which the unfolding was performed with a fine $z$ but coarse $x$ binning (see Table~\ref{tab:binning}). 
The low-$z$ bin spans the range $0.1<z<0.2$, which is excluded from asymmetries that are integrated over $z$.
The resulting $A_{\parallel}^h(z)$ is shown for the three $x$ slices in Fig.~\ref{fig:2DXZ}. No strong dependence on $z$ is visible, in agreement with results by the COMPASS collaboration for charged-hadron production from longitudinally polarized deuterons~\cite{Alekseev:2010dm,Adolph:2016vou}.

To better evaluate any potential $z$ dependence, and in order to avoid, e.g., possible influence of the $y$ dependence of $A^{h}_{\parallel}$ through its kinematic prefactors, $A^{h}_{1}$  was determined from $A^{h}_{\parallel}$ according to Eq.~\eqref{eq:A1exp}. 
A set of polynomial functions---one linear in $x$ only, one linear in both $x$ and $z$, 
and one second order in both variables---was then fit to all 18 data points with correlated uncertainties for each of the resulting $A^{h}_{1}$ asymmetries.
It was found that within the 
precision of the asymmetries, the goodness-of-fit was not significantly improved by 
including a $z$ dependence. The $\chi^2$ values are given in Table~\ref{tab:xz-chisq}.

\begin{figure*}
	\centering
	\includegraphics[width=0.67\textwidth]{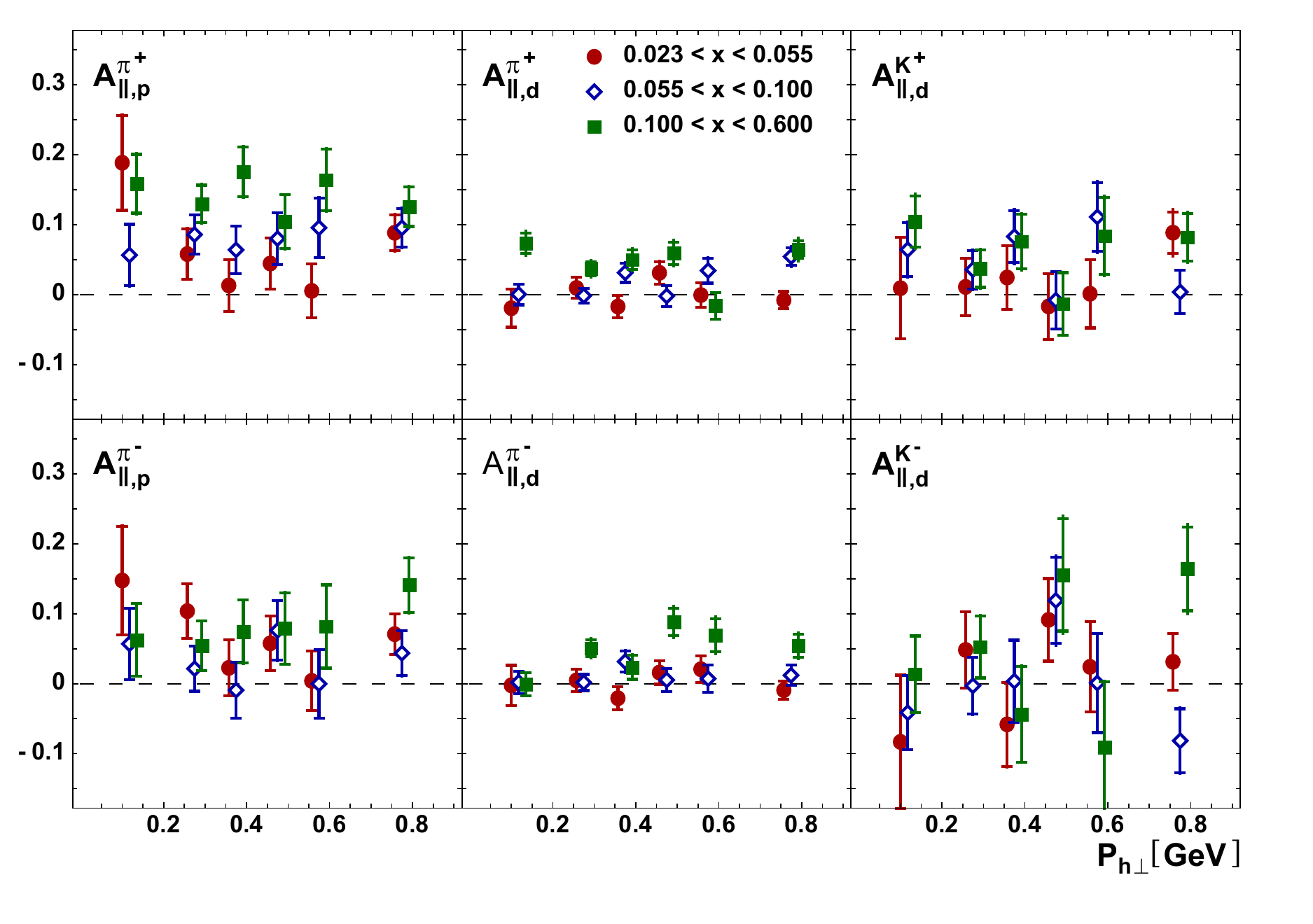}
	\caption{   The longitudinal double-spin asymmetries $A_{\parallel, N}^h$ as a function of $P_{h\perp}$ 
	                 in three different $x$ ranges as labeled, 
			with $N=p,d$ denoting the target nucleus and $h=\pi^{\pm},K^{\pm}$ the final-state hadron detected.
            Data points for the first $x$ slice are plotted at their average kinematics, while the ones for 
			the other two $x$ slices are slightly shifted horizontally for better legibility.
			The inner error bars represent statistical uncertainties while the outer ones 
			statistical and systematic uncertainties added in quadrature.
			}
	\label{fig:A1xpt}
\end{figure*}

\begin{table*}
     \begin{ruledtabular}
	\caption{
		The $\chi^2$ values for polynomial fits to the $A^{h}_{1,N}(x,P_{h\perp})$ 
		data points for each combination of target (\(N=p,d\)) and final-state hadron \(h\), 
			and number of degrees of freedom as indicated. 
            The $C_i^h$ are the polynomial terms of the fit functions.
		The fit function linear in $P_{h\perp}$ yields little improvement over the fit 
		constant in that variable suggesting little or no $P_{h\perp}$ 
		dependence of the asymmetry within the 
		statistical precision of the data. 
			}
	\label{tab:chisq}
	\begin{tabular}{lcccccc}
		~& \( A^{\pi^{+}}_{1,p} \) & \( A^{\pi^{-}}_{1,p} \) & \( A^{\pi^{+}}_{1,d} \) & \( A^{\pi^{-}}_{1,d} \) & \( A^{K^{+}}_{1,d} \) & \( A^{K^{-}}_{1,d} \)   \\[.5mm]
		\hline 
		$\chi^{2~(NDF=16)}_{~C_1^h+C_2^hx}$ & 12.7 & 14.0 & 33.7 & 22.9 & 16.0 & 24.4 \\[3mm] 
		$\chi^{2~(NDF=15)}_{~C_1^h+C_2^hx+C_3^hP_{h\perp}}$ & 12.7 & 13.9 & 31.9 & 20.6 & 16.0 & 23.6 \\[3mm] 
		$\chi^{2~(NDF=12)}_{~C_1^h+C_2^hx+C_3^hP_{h\perp}+C_{4}^{h} x^{2} + C_{5}^{h} P_{h\perp}^{2} +C_{6}^{h} xP_{h\perp}}$ & 8.5&5.1&29.7&12.0&12.2&18.7\\[.5mm]
	\end{tabular}
     \end{ruledtabular}
\end{table*}

\begin{figure}
	\centering
	\includegraphics[width= 0.35\textwidth]{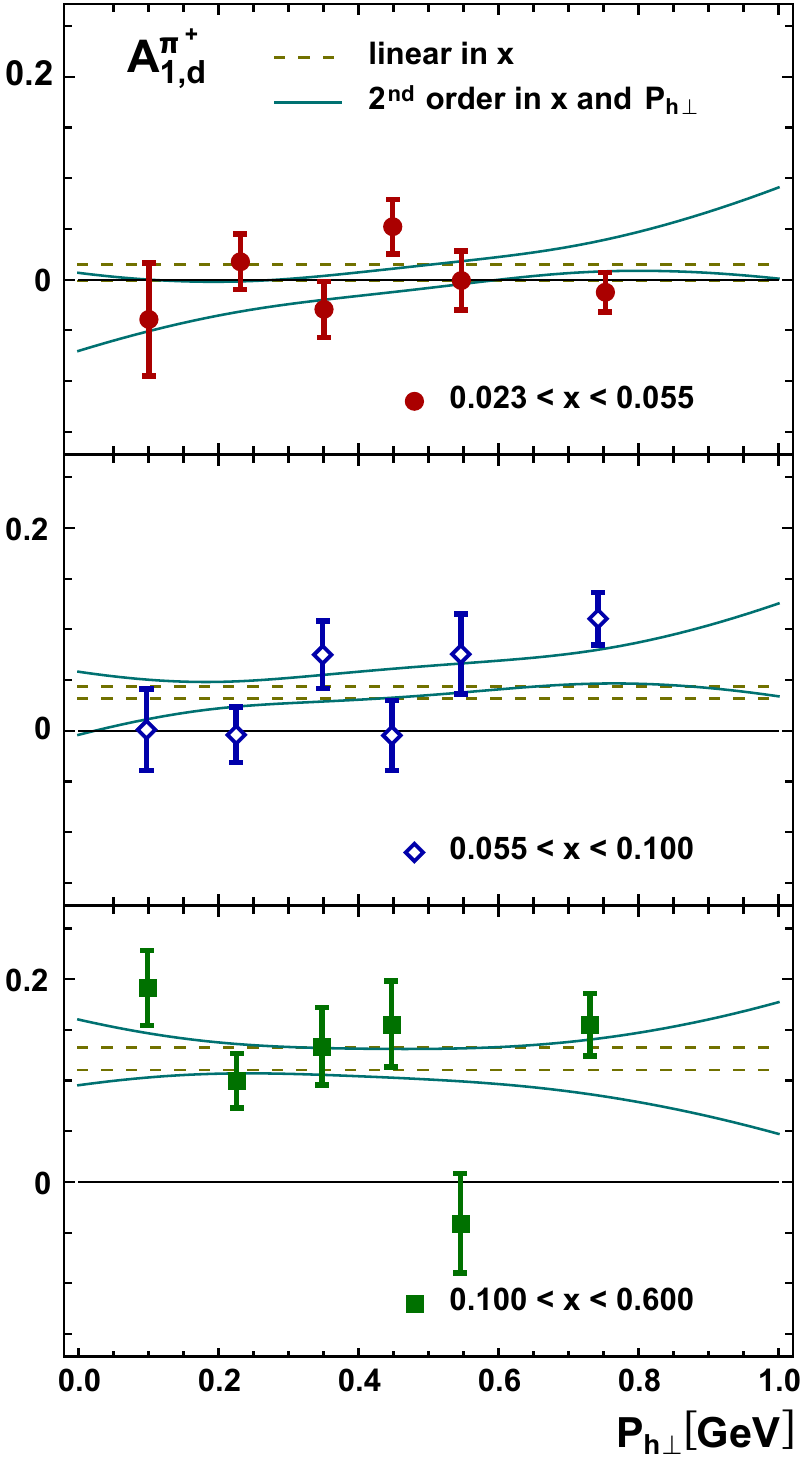}
	\caption{$A_{1,d}^{\pi^+}$    
			shown for three separate ranges in $x$
			with the \(1\sigma\) uncertainty bands of two analytic fits. 
            One fit is linear in $x$ only (dashed line) and 
			one is a second-order polynomial in both $x$ and $P_{h\perp}$ (full line).
			These fits are intended to convey the statistical significance of the dataset 
			which includes significant bin-to-bin correlations.
			As can be seen by the $\chi^2$ values in Table~\ref{tab:chisq} 
            the data do not favor any of the functional forms studied.
			}
	\label{fig:A1xpt-fitcol}
\end{figure}

The $x$-$P_{h\perp}$ dependence of $A^{h}_{\parallel}$ is obtained by 
binning and unfolding in both of these variables simultaneously 
(see Table~\ref{tab:binning}), as done for the $x$-$z$ projection of $A^{h}_{\parallel}$.	  
A dependence on the transverse hadron momentum may arise from different 
average transverse momenta of quarks with their spin aligned to the nucleon spin compared to the case of the spins being anti-aligned. 
The asymmetries are presented in Fig.~\ref{fig:A1xpt} as a function of $P_{h\perp}$ for three disjoint $x$ ranges. No strong dependence on $P_{h\perp}$ is visible, consistent with the weak dependences reported by the CLAS~\cite{Avakian:2010ae} and COMPASS~\cite{Alekseev:2010dm,Adolph:2016vou} collaborations.

	In order to evaluate in more detail any potential $P_{h\perp}$ dependence, each of the asymmetries 
	was transformed into a corresponding $A^{h}_{1}$ asymmetry and then fit with a set of polynomial functions as  
	was done for the $x$--$z$ dependence---one linear in $x$ only, one 
	linear in both $x$ and $P_{h\perp}$, and second order in both variables. 
	Again, the goodness-of-fit of these polynomial fit functions, given in Table~\ref{tab:chisq}, 
	shows no clear preference for any of the functional forms used.
	Figure~\ref{fig:A1xpt-fitcol} shows as an example $A_{1}^{\pi^+}(P_{h\perp})$ from deuterons in three  
	$x$ ranges as given in the different panels. Uncertainty bands are overlaid for two fits.  
	They are presented to provide 
	a realistic indication of the model-constraining power of these data.

\subsection{The semi-inclusive asymmetry binned in three dimensions}\label{sec:3dresults}

\begin{figure}[h]
	\centering
	\includegraphics[width=0.48\textwidth]{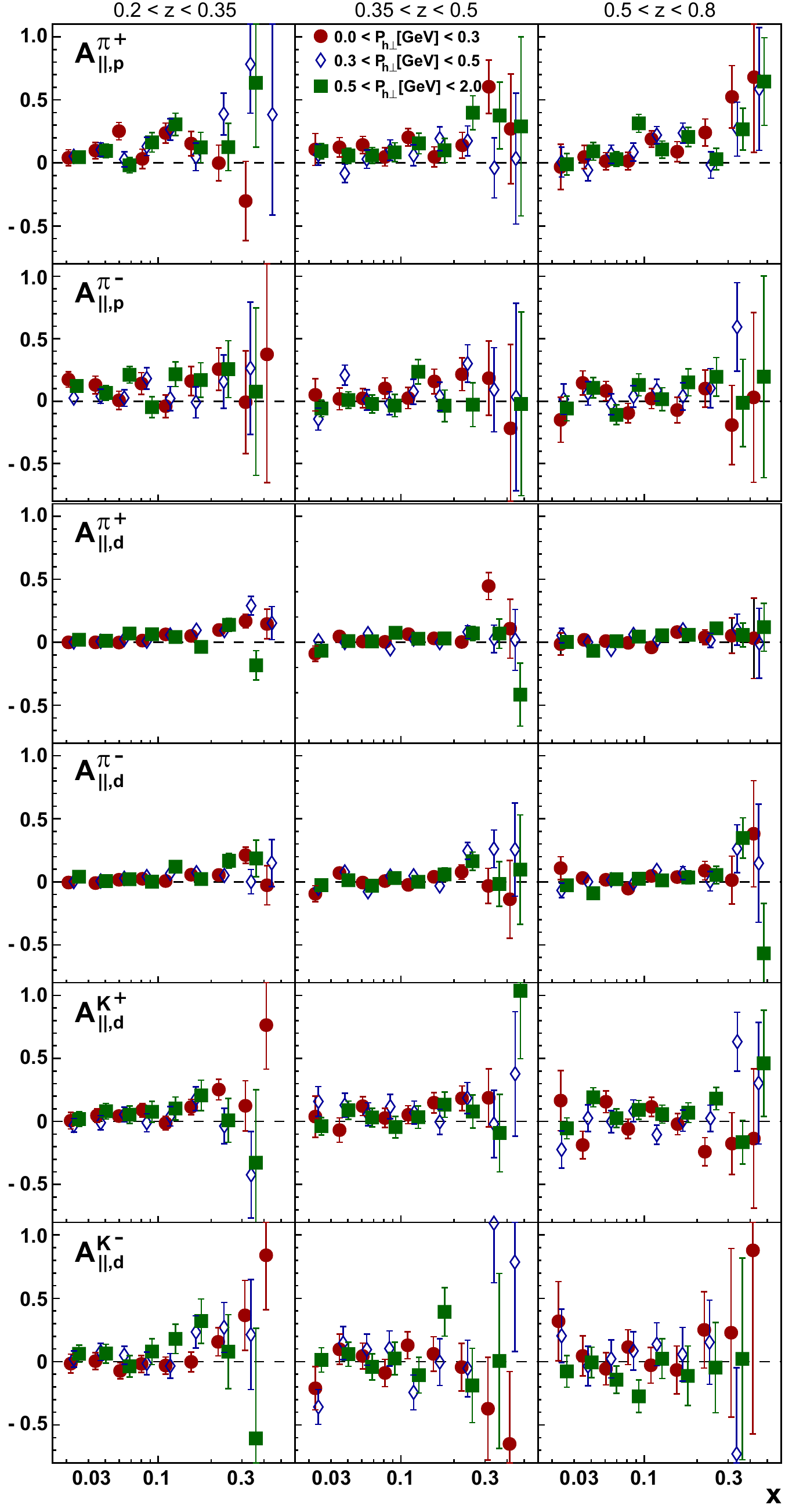}
	\caption{   $A_{\parallel, N}^h(x,z,P_{h\perp})$ as a function of \(x\) in three different $z$ ranges and three 
	                 different $P_{h\perp}$ ranges as labeled (see Table~\ref{tab:binning} for details), 
			with $N=p,d$ denoting the target nucleus and $h=\pi^{\pm},K^{\pm}$ the final-state hadron detected.
			Data points for the second $P_{h\perp}$ slice are plotted at their average kinematics, while the ones for 
			the remaining $P_{h\perp}$ slices are slightly shifted horizontally for better legibility.
            The inner error bars represent statistical uncertainties while the outer ones statistical and systematic uncertainties added in quadrature.
			}
	\label{fig:Apar-p-xzpt}
\end{figure}

	The hadron-tagged longitudinal double-spin asymmetry binned simultaneously in $x$, $z$, and 
	$P_{h\perp}$ as measured by HERMES for hydrogen and deuterium targets are 
	presented in Figs.~\ref{fig:Apar-p-xzpt}.  
	The asymmetry is binned in a grid with nine bins in $x$, 
	three bins in $P_{h\perp}$, 
	and three bins in $z$ 
	(see Table~\ref{tab:binning}), and is plotted as a function of \(x\) for those ranges in \(z\) and \(P_{h\perp}\).
	The binning was selected to populate the bins with statistics as uniformly as 
	reasonable while maintaining a degree of kinematic uniformity across each bin. 
	Within the precision of the measurements, the asymmetries display no obvious dependence on the hadron variables. 
	There is possibly an indication that the non-vanishing asymmetry for \(\pi^{-}\) from protons observed in the one-dimensional binning in \(x\)  (cf.~Fig.~\ref{fig:lowQ2d})
	is caused to a large extent by low-\(z\) pions. This is in line with expectation considering that disfavored fragmentation, e.g., fragmentation of quark flavors 
	that are not part of the valence structure of the hadron produced, is sizable in that region. As such, \(\pi^{-}\) production from up quarks---which 
	carry a large positive asymmetry---may still play a dominant role in that kinematic region compared
	to larger values of \(z\), where disfavored fragmentation will be more and more suppressed.

	These data as well as those of the other asymmetry results discussed are available 
	as Supplemental Material \cite{asymData}. 
	A statistical covariance matrix is also provided, which 
	describes the uncertainties of the asymmetry in every kinematic bin as well as the degree 
	of correlation between them, which comes about as a result of the 
	unfolding process. This complete covariance information should be 
	included in any derivative calculation as omitting it---that is using the 
	single-bin uncertainties alone---underestimates the statistical significance of 
	these data. 
	These three-dimensionally binned asymmetries are the most complete, unintegrated, longitudinally polarized double-spin dataset to date.

\subsection{Azimuthal asymmetries}\label{sec:azimresults}

	\begin{figure*}[t]
		\centering
		\includegraphics[width=0.69 \textwidth]{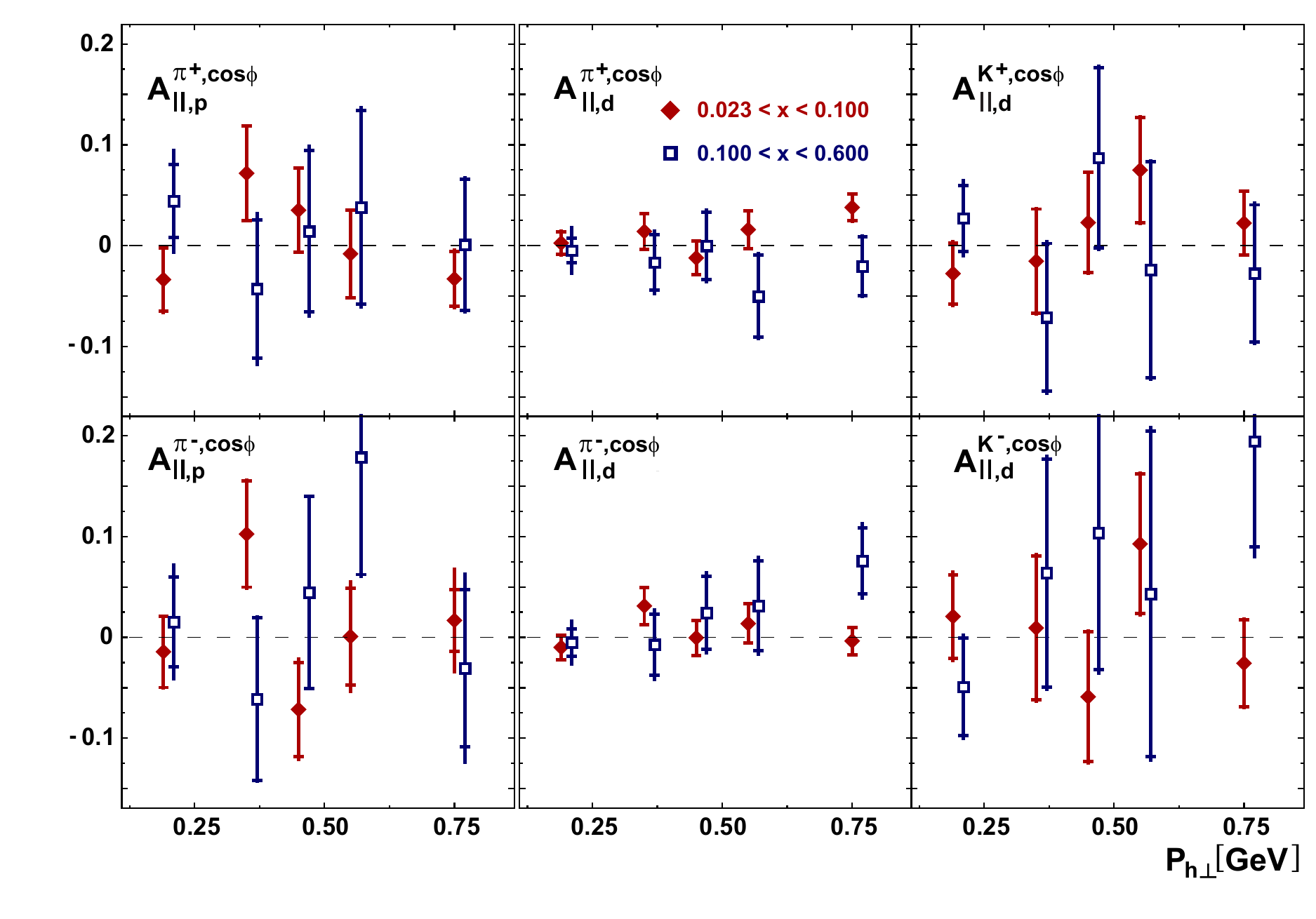}
		\caption{$A_{\parallel}^{h,\cos\phi}(P_{h\perp})$ in two $x$ ranges for charged pions (and kaons) from protons (deuterons) as labelled.
			The inner error bars represent statistical uncertainties while the outer ones 
			statistical and systematic uncertainties added in quadrature. 
			Data points for the first $x$ slice are plotted at their average kinematics, while the ones for 
			the second $x$ slice are slightly shifted horizontally for better legibility.
		}
		\label{fig:ALLpt}
	\end{figure*}

As described in the introduction, azimuthal moments of asymmetries are potentially sensitive to unique combinations of distribution and fragmentation functions, a number of which vanish when integrated over semi-inclusive kinematic parameters.

\begin{table}[b]
     \begin{ruledtabular}
	\caption{
		Bin boundaries used for the various projections of  \(A_{LL}^{h,\cos\phi}\).}
	\label{tab:ALLcosphiBinning}
	\begin{tabular}{cc}
	     \(x\) binning   & \(z\) binning \\
	\hline     
	0.023 -- 0.1 -- 0.6 & 0.2 -- 0.32 -- 0.44 -- 0.56 --0.68 -- 0.8 \\
	\hline\hline\\[1mm]
	\hline\hline
	     \(x\) binning   & \(P_{h\perp}\)[ GeV] binning   \\
	\hline     
	0.023 -- 0.1 -- 0.6 & 0 -- 0.3 -- 0.4 -- 0.5 --0.6 -- 2 	 \\
	\hline\hline\\[1mm]
	\hline\hline
	    \(z\) binning   & \(x\) binning \\
	\hline
	0.2 -- 0.4 -- 0.6 & 0.023 -- 0.04 -- 0.055 -- 0.075 -- 0.14 -- 0.6 
		\end{tabular}
     \end{ruledtabular}
\end{table}

For each hadron and target combination, the asymmetry is divided into 10 $\phi$ bins and fit with an azimuthally periodic function in each of either 2 $x$ $\times$  5 $z$-bins, 2 $x$ $\times$  5 $P_{h\perp}$-bins, or  2 $z$ $\times$  5 $x$-bins as detailed in Table~\ref{tab:ALLcosphiBinning}.  
The functional form used included constant, $\cos\phi$, and $\cos 2\phi$ terms.
Each of these cosine moments is found to be consistent with zero. 
(A similar result was obtained for unidentified hadrons for deuteron data from the COMPASS experiment~\cite{Alekseev:2010dm,Adolph:2016vou}.)
The $P_{h\perp}$ projections of the $\cos \phi$ moments for charged pions for each target, as well as for charged kaons in case of a deuterium target are presented in Fig.~\ref{fig:ALLpt}. All other projections of the $\cos \phi$ moments are included in the data tables in~\cite{asymData}, including the statistically more precise results for unidentified hadrons.\footnote{Note that here and in the later discussed hadron charge-difference asymmetry the momentum requirement for unidentified hadrons is relaxed to \(P_{h}>0.5\)~GeV.}
 
A vanishing $\cos 2\phi$ asymmetry as found here can be expected because in the one-photon-exchange approximation there is no $A_{LL}^{h,\cos 2\phi}$ contribution to the cross section [cf.~Eq.~\eqref{eqn:big-cross}] and thus a non-zero $A_{\parallel}^{h,\cos2\phi}$ can arise in this approximation only through the very small transverse component of the target-spin vector in a configuration where the target is polarized along the beam direction~\cite{Diehl:2005}.

 \subsection{The hadron charge-difference asymmetry}\label{sec:charge-difference}

\begin{figure}[b]
		\centering
		\includegraphics[width=0.4 \textwidth]{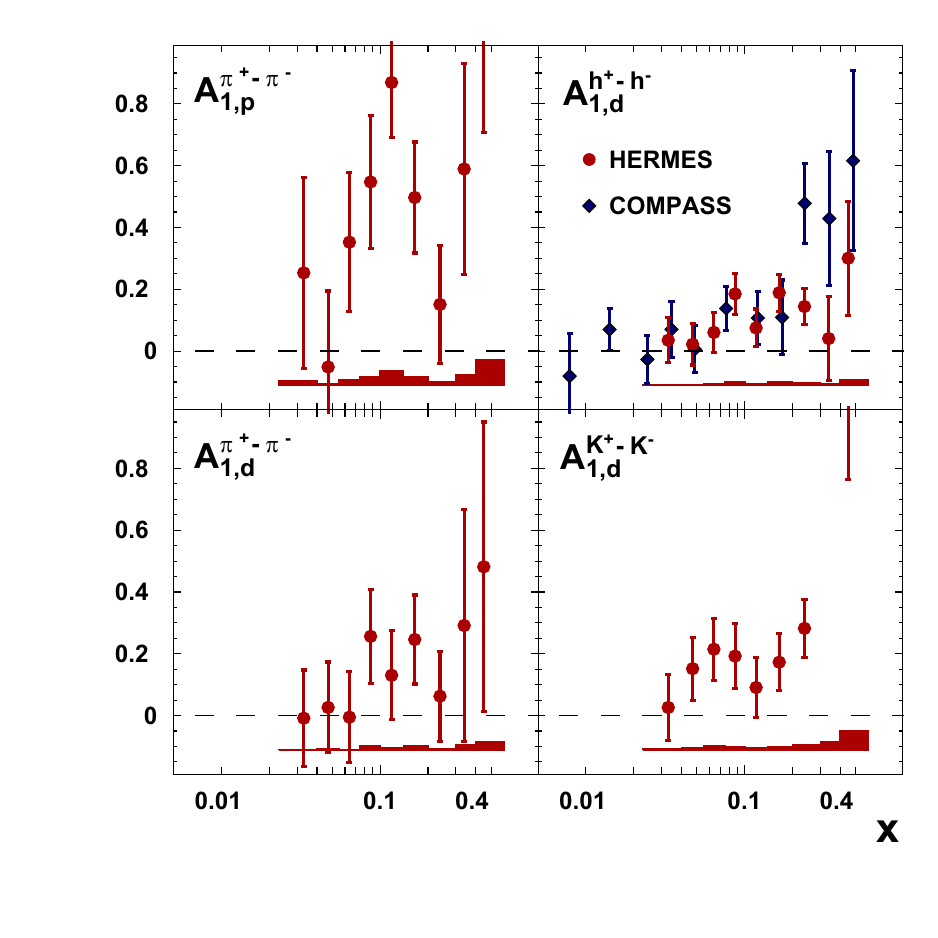}
		\caption{
			Hadron charge-difference asymmetries for pions from the hydrogen target and pions, 
			kaons, and all hadrons from the deuterium target. Error bars represent statistical uncertainties. Systematic uncertainties are given as bands. Data from COMPASS \cite{Alekseev:2007vi} for undifferentiated hadrons using a $^{6}$LiD target are also shown.
		}
		\label{fig:diffAsy}
	\end{figure}

The hadron charge-difference asymmetry
	\begin{equation}
		A_1^{h^+ - h^-}(x) \equiv 
		\frac{\left(\sigma_{1/2}^{h^+}-\sigma_{1/2}^{h^-}\right)-\left(\sigma_{3 
		/2}^{h^+}-\sigma_{3/2}^{h^-}\right)}{\left(\sigma_{1/2}^{h^+}-\sigma_{1/ 
		2}^{h^-}\right)+\left(\sigma_{3/2}^{h^+}-\sigma_{3/2}^{h^-}\right)}
		\label{eq:A1ChargeDiff}
	\end{equation}
	provides additional spin-structure information and is not trivially 
	constructible from the simple semi-inclusive asymmetries.  The difference asymmetries
	for pions from the hydrogen target and pions, kaons, and undifferentiated hadrons from the deuterium target
	are shown in Fig.~\ref{fig:diffAsy},
    together with results from the COMPASS Collaboration for unidentified hadrons from a $^6$LiD target~\cite{Alekseev:2007vi}. 
	A feature that might be unexpected is that the uncertainties for the 
	kaon asymmetry are considerably smaller than those on the pion asymmetry despite the 
	smaller sample size.  This is a result of the larger difference between yields of charged 
	kaons compared to that of the charged pions (as $K^-$ shares no valence quarks with 
	the target), which causes a significantly larger denominator of Eq.~\eqref{eq:A1ChargeDiff}.
	
	Under the assumption of leading-order (LO), leading-twist (LT)
	QCD,  and charge-conjugation symmetry of the fragmentation functions, i.e.,
	\begin{equation}
		D_1^{q\to h^+}=D_1^{\bar{q} \to h^-},
		\label{eq:symAssume}
	\end{equation}
	the difference asymmetry on the deuteron may be equated to a certain combination of parton distributions~\cite{Frankfurt:1989wq}:
	\begin{equation}
		  A_{1,d}^{h^+ - h^-}   \stackrel{\textrm{\tiny{LO LT}}}{=} 
		\frac{g_{1}^{u_v}+g_{1}^{d_v}}{f_{1}^{u_v}+f_{1}^{d_v}} .
		\label{eq:diff-deuteron}
		\end{equation}
	Here, $f_{1}^{q_v} \equiv f_{1}^{q}-f_{1}^{\bar{q}} $ ($g_{1}^{q_v} \equiv g_{1}^{q}-g_{1}^{\bar{q}} $) is the  polarization-averaged (helicity) valence-quark distribution of the proton, and ``LO LT" is a reminder
	of the assumptions mentioned previously.  This is equivalent to assuming a well differentiated current and target region; 
	a scenario in which the struck quark has no memory of the hadron variety to which it previously belonged.
	
	By further assuming isospin symmetry in fragmentation, that is
	\begin{equation}
		D_1^{u \to \pi^+}=D_{1}^{d \to \pi^-}~~\text{and}~~D_1^{u \to \pi^-}=D_{1}^{d\to\pi^+} ,	
	\end{equation}
	a second valence-quark expression using charge-difference asymmetries from a hydrogen target is given by
		\begin{equation}
		 A_{1,p}^{h^+ - h^-}  \stackrel{\textrm{\tiny{LO LT}}}{=} 
		\frac{4 g_{1}^{u_v}-g_{1}^{d_v}}{4 f_{1}^{u_v}-f_{1}^{d_v}} \; .
		\label{eq:diff-proton}
	\end{equation}

	\begin{figure}
		\centering
		\includegraphics[width=0.42 \textwidth]{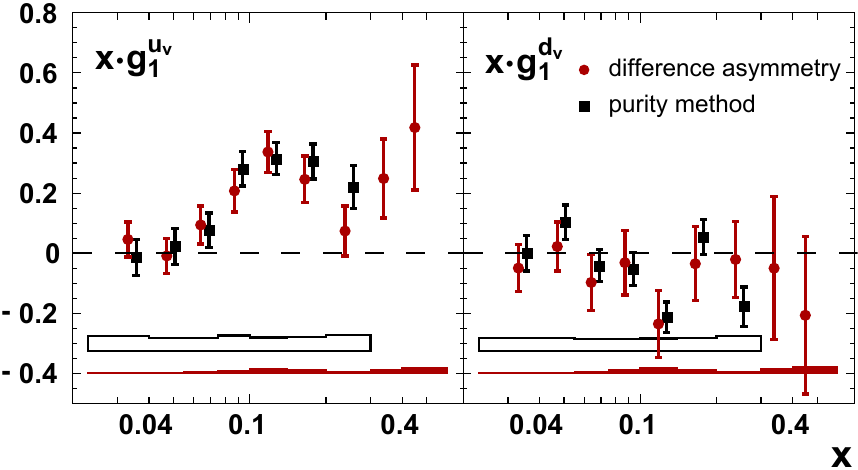}
		\caption{
			Helicity distributions for valence quarks computed using pion charge-difference asymmetries
			and Eqs.~\eqref{eq:diff-deuteron} and \eqref{eq:diff-proton} compared with
			valence-quark densities (as indicated) computed from the HERMES purity extraction
			\cite{Airapetian:2004zf}. Error bars represent statistical uncertainties.
            	Systematic uncertainties from the difference-asymmetry (purity) extraction are shown as filled (open) bands.
		}
		\label{fig:helicityDensities}
	\end{figure}

It follows that the charge-difference asymmetries should be independent of the
hadron type, a feature consistent with the results shown in Fig.~\ref{fig:diffAsy}.  Valence-quark helicity densities computed using Eqs.~\eqref{eq:diff-deuteron} and \eqref{eq:diff-proton} are presented in Fig.~\ref{fig:helicityDensities} alongside the same quantities computed from the previous HERMES purity extraction~\cite{Airapetian:2004zf}.  The results are largely consistent using two methods that have very different and quite complementary model assumptions.  Whereas the method presented here depends on leading-order and leading-twist assumptions to provide the clean factorization, which ensures that fragmentation can proceed without memory of the target configuration, the purity method depends on a fragmentation model subject to its own uncertainties related to the model tune and the believability of its phenomenologically motivated dynamics. The lack of dependence on hadron type of the charge-difference asymmetries and the consistency of the derived valence-quark helicity distributions with the results of the purity analysis suggest that there is no significant deviation from the factorization hypothesis.

\section{Conclusion}

Several longitudinal double-spin asymmetries in semi-inclusive deep-inelastic scattering have been presented. 
They extend the analysis of the previous HERMES publications to include also transverse-momentum dependence.
Within the precision of the measurements, the virtual-photon--nucleon asymmetries $A_{1}^h(x,z)$ and $A_{1}^h(x, P_{h\perp})$ display no significant dependence on the hadron variables. 
Azimuthal moments, $A_{\parallel}^{h,\cos\phi}$, are found to be consistent with zero.
The hadron charge-difference asymmetry $A_1^{h^+-h^-}(x)$ yields valence-quark helicity densities consistent with the result of the prior HERMES purity extraction.  
A common thread among these results is that within the available statistical precision the longitudinal sector shows no deviation from the leading-order, leading-twist assumption.  
In addition to this interpretation, these data are expected to provide an essentially model-independent constraint for theory and parameterization as they provide the first ever longitudinal double-spin semi-inclusive dataset binned in as many as three kinematic variables simultaneously.
They point the way to future precision tests of models of nucleon structure that go beyond a collinear framework.

\acknowledgments

We gratefully acknowledge the DESY management for its support, the staff
at DESY and the collaborating institutions for their significant effort.
This work was supported by 
the State Committee of Science of the Republic of Armenia, Grant No. 18T-1C180;
the FWO-Flanders and IWT, Belgium;
the Natural Sciences and Engineering Research Council of Canada;
the National Natural Science Foundation of China;
the Alexander von Humboldt Stiftung,
the German Bundesministerium f\"ur Bildung und Forschung (BMBF), and
the Deutsche Forschungsgemeinschaft (DFG);
the Italian Istituto Nazionale di Fisica Nucleare (INFN);
the MEXT, JSPS, and G-COE of Japan;
the Dutch Foundation for Fundamenteel Onderzoek der Materie (FOM);
the Russian Academy of Science and the Russian Federal Agency for 
Science and Innovations;
the Basque Foundation for Science (IKERBASQUE), the Basque Government, Grant No. IT956-16, and MINECO (Juan de la Cierva), Spain;
the U.K.~Engineering and Physical Sciences Research Council, 
the Science and Technology Facilities Council,
and the Scottish Universities Physics Alliance;
as well as the U.S.~Department of Energy (DOE) and the National Science Foundation (NSF).

\bibliographystyle{apsrev4-1}
\bibliography{DC85}

\begin{thebibliography}{43}%
\makeatletter
\providecommand \@ifxundefined [1]{%
 \@ifx{#1\undefined}
}%
\providecommand \@ifnum [1]{%
 \ifnum #1\expandafter \@firstoftwo
 \else \expandafter \@secondoftwo
 \fi
}%
\providecommand \@ifx [1]{%
 \ifx #1\expandafter \@firstoftwo
 \else \expandafter \@secondoftwo
 \fi
}%
\providecommand \natexlab [1]{#1}%
\providecommand \enquote  [1]{``#1''}%
\providecommand \bibnamefont  [1]{#1}%
\providecommand \bibfnamefont [1]{#1}%
\providecommand \citenamefont [1]{#1}%
\providecommand \href@noop [0]{\@secondoftwo}%
\providecommand \href [0]{\begingroup \@sanitize@url \@href}%
\providecommand \@href[1]{\@@startlink{#1}\@@href}%
\providecommand \@@href[1]{\endgroup#1\@@endlink}%
\providecommand \@sanitize@url [0]{\catcode `\\12\catcode `\$12\catcode
  `\&12\catcode `\#12\catcode `\^12\catcode `\_12\catcode `\%12\relax}%
\providecommand \@@startlink[1]{}%
\providecommand \@@endlink[0]{}%
\providecommand \url  [0]{\begingroup\@sanitize@url \@url }%
\providecommand \@url [1]{\endgroup\@href {#1}{\urlprefix }}%
\providecommand \urlprefix  [0]{URL }%
\providecommand \Eprint [0]{\href }%
\providecommand \doibase [0]{http://dx.doi.org/}%
\providecommand \selectlanguage [0]{\@gobble}%
\providecommand \bibinfo  [0]{\@secondoftwo}%
\providecommand \bibfield  [0]{\@secondoftwo}%
\providecommand \translation [1]{[#1]}%
\providecommand \BibitemOpen [0]{}%
\providecommand \bibitemStop [0]{}%
\providecommand \bibitemNoStop [0]{.\EOS\space}%
\providecommand \EOS [0]{\spacefactor3000\relax}%
\providecommand \BibitemShut  [1]{\csname bibitem#1\endcsname}%
\let\auto@bib@innerbib\@empty
\bibitem [{\citenamefont {Aidala}\ \emph {et~al.}(2013)\citenamefont {Aidala},
  \citenamefont {Bass}, \citenamefont {Hasch},\ and\ \citenamefont
  {Mallot}}]{Aidala:2012mv}%
  \BibitemOpen
  \bibfield  {author} {\bibinfo {author} {\bibfnamefont {C.~A.}\ \bibnamefont
  {Aidala}}, \bibinfo {author} {\bibfnamefont {S.~D.}\ \bibnamefont {Bass}},
  \bibinfo {author} {\bibfnamefont {D.}~\bibnamefont {Hasch}}, \ and\ \bibinfo
  {author} {\bibfnamefont {G.~K.}\ \bibnamefont {Mallot}},\ }\href {\doibase
  10.1103/RevModPhys.85.655} {\bibfield  {journal} {\bibinfo  {journal} {Rev.
  Mod. Phys.}\ }\textbf {\bibinfo {volume} {85}},\ \bibinfo {pages} {655}
  (\bibinfo {year} {2013})},\ \Eprint {http://arxiv.org/abs/1209.2803}
  {arXiv:1209.2803 [hep-ph]} \BibitemShut {NoStop}%
\bibitem [{\citenamefont {Adeva}\ \emph {et~al.}(1998)\citenamefont {Adeva}
  \emph {et~al.}}]{Adeva:1997qz}%
  \BibitemOpen
  \bibfield  {author} {\bibinfo {author} {\bibfnamefont {B.}~\bibnamefont
  {Adeva}} \emph {et~al.} (\bibinfo {collaboration} {Spin Muon
  Collaboration}),\ }\href {\doibase 10.1016/S0370-2693(97)01546-3} {\bibfield
  {journal} {\bibinfo  {journal} {Phys. Lett.}\ }\textbf {\bibinfo {volume}
  {B420}},\ \bibinfo {pages} {180} (\bibinfo {year} {1998})},\ \Eprint
  {http://arxiv.org/abs/hep-ex/9711008} {hep-ex/9711008} \BibitemShut {NoStop}%
\bibitem [{\citenamefont {Ackerstaff}\ \emph {et~al.}(1999)\citenamefont
  {Ackerstaff} \emph {et~al.}}]{Ackerstaff:1999ey}%
  \BibitemOpen
  \bibfield  {author} {\bibinfo {author} {\bibfnamefont {K.}~\bibnamefont
  {Ackerstaff}} \emph {et~al.} (\bibinfo {collaboration} {HERMES
  Collaboration}),\ }\href {\doibase 10.1016/S0370-2693(99)00964-8} {\bibfield
  {journal} {\bibinfo  {journal} {Phys. Lett.}\ }\textbf {\bibinfo {volume}
  {B464}},\ \bibinfo {pages} {123} (\bibinfo {year} {1999})},\ \Eprint
  {http://arxiv.org/abs/hep-ex/9906035} {hep-ex/9906035} \BibitemShut {NoStop}%
\bibitem [{\citenamefont {Airapetian}\ \emph {et~al.}(2004)\citenamefont
  {Airapetian} \emph {et~al.}}]{Airapetian:2003ct}%
  \BibitemOpen
  \bibfield  {author} {\bibinfo {author} {\bibfnamefont {A.}~\bibnamefont
  {Airapetian}} \emph {et~al.} (\bibinfo {collaboration} {HERMES
  Collaboration}),\ }\href {\doibase 10.1103/PhysRevLett.92.012005} {\bibfield
  {journal} {\bibinfo  {journal} {Phys. Rev. Lett.}\ }\textbf {\bibinfo
  {volume} {92}},\ \bibinfo {pages} {012005} (\bibinfo {year} {2004})},\
  \Eprint {http://arxiv.org/abs/hep-ex/0307064} {hep-ex/0307064} \BibitemShut
  {NoStop}%
\bibitem [{\citenamefont {Airapetian}\ \emph
  {et~al.}(2005{\natexlab{a}})\citenamefont {Airapetian} \emph
  {et~al.}}]{Airapetian:2004zf}%
  \BibitemOpen
  \bibfield  {author} {\bibinfo {author} {\bibfnamefont {A.}~\bibnamefont
  {Airapetian}} \emph {et~al.} (\bibinfo {collaboration} {HERMES
  Collaboration}),\ }\href {\doibase 10.1103/PhysRevD.71.012003} {\bibfield
  {journal} {\bibinfo  {journal} {Phys. Rev.}\ }\textbf {\bibinfo {volume}
  {D71}},\ \bibinfo {pages} {012003} (\bibinfo {year} {2005}{\natexlab{a}})},\
  \Eprint {http://arxiv.org/abs/hep-ex/0407032} {hep-ex/0407032} \BibitemShut
  {NoStop}%
\bibitem [{\citenamefont {Alekseev}\ \emph {et~al.}(2008)\citenamefont
  {Alekseev} \emph {et~al.}}]{Alekseev:2007vi}%
  \BibitemOpen
  \bibfield  {author} {\bibinfo {author} {\bibfnamefont {M.}~\bibnamefont
  {Alekseev}} \emph {et~al.} (\bibinfo {collaboration} {COMPASS
  Collaboration}),\ }\href {\doibase 10.1016/j.physletb.2007.12.056} {\bibfield
   {journal} {\bibinfo  {journal} {Phys. Lett.}\ }\textbf {\bibinfo {volume}
  {B660}},\ \bibinfo {pages} {458} (\bibinfo {year} {2008})},\ \Eprint
  {http://arxiv.org/abs/0707.4077} {arXiv:0707.4077 [hep-ex]} \BibitemShut
  {NoStop}%
\bibitem [{\citenamefont {Airapetian}\ \emph {et~al.}(2008)\citenamefont
  {Airapetian} \emph {et~al.}}]{Airapetian:2008qf}%
  \BibitemOpen
  \bibfield  {author} {\bibinfo {author} {\bibfnamefont {A.}~\bibnamefont
  {Airapetian}} \emph {et~al.} (\bibinfo {collaboration} {HERMES
  Collaboration}),\ }\href {\doibase 10.1016/j.physletb.2008.07.090} {\bibfield
   {journal} {\bibinfo  {journal} {Phys. Lett.}\ }\textbf {\bibinfo {volume}
  {B666}},\ \bibinfo {pages} {446} (\bibinfo {year} {2008})},\ \Eprint
  {http://arxiv.org/abs/0803.2993} {arXiv:0803.2993 [hep-ex]} \BibitemShut
  {NoStop}%
\bibitem [{\citenamefont {Alekseev}\ \emph {et~al.}(2009)\citenamefont
  {Alekseev} \emph {et~al.}}]{Alekseev:2009ac}%
  \BibitemOpen
  \bibfield  {author} {\bibinfo {author} {\bibfnamefont {M.}~\bibnamefont
  {Alekseev}} \emph {et~al.} (\bibinfo {collaboration} {COMPASS
  Collaboration}),\ }\href {\doibase 10.1016/j.physletb.2009.08.065} {\bibfield
   {journal} {\bibinfo  {journal} {Phys. Lett.}\ }\textbf {\bibinfo {volume}
  {B680}},\ \bibinfo {pages} {217} (\bibinfo {year} {2009})},\ \Eprint
  {http://arxiv.org/abs/0905.2828} {arXiv:0905.2828 [hep-ex]} \BibitemShut
  {NoStop}%
\bibitem [{\citenamefont {Alekseev}\ \emph
  {et~al.}(2010{\natexlab{a}})\citenamefont {Alekseev} \emph
  {et~al.}}]{Alekseev:2010ub}%
  \BibitemOpen
  \bibfield  {author} {\bibinfo {author} {\bibfnamefont {M.}~\bibnamefont
  {Alekseev}} \emph {et~al.} (\bibinfo {collaboration} {COMPASS
  Collaboration}),\ }\href {\doibase 10.1016/j.physletb.2010.08.034} {\bibfield
   {journal} {\bibinfo  {journal} {Phys. Lett.}\ }\textbf {\bibinfo {volume}
  {B693}},\ \bibinfo {pages} {227} (\bibinfo {year} {2010}{\natexlab{a}})},\
  \Eprint {http://arxiv.org/abs/1007.4061} {arXiv:1007.4061 [hep-ex]}
  \BibitemShut {NoStop}%
\bibitem [{\citenamefont {Avakian}\ \emph {et~al.}(2010)\citenamefont {Avakian}
  \emph {et~al.}}]{Avakian:2010ae}%
  \BibitemOpen
  \bibfield  {author} {\bibinfo {author} {\bibfnamefont {H.}~\bibnamefont
  {Avakian}} \emph {et~al.} (\bibinfo {collaboration} {CLAS Collaboration}),\
  }\href {\doibase 10.1103/PhysRevLett.105.262002} {\bibfield  {journal}
  {\bibinfo  {journal} {Phys. Rev. Lett.}\ }\textbf {\bibinfo {volume} {105}},\
  \bibinfo {pages} {262002} (\bibinfo {year} {2010})},\ \Eprint
  {http://arxiv.org/abs/1003.4549} {arXiv:1003.4549 [hep-ex]} \BibitemShut
  {NoStop}%
\bibitem [{\citenamefont {Jawalkar}\ \emph {et~al.}(2018)\citenamefont
  {Jawalkar} \emph {et~al.}}]{Jawalkar:2017ube}%
  \BibitemOpen
  \bibfield  {author} {\bibinfo {author} {\bibfnamefont {S.}~\bibnamefont
  {Jawalkar}} \emph {et~al.} (\bibinfo {collaboration} {CLAS Collaboration}),\
  }\href {\doibase 10.1016/j.physletb.2018.06.014} {\bibfield  {journal}
  {\bibinfo  {journal} {Phys. Lett.}\ }\textbf {\bibinfo {volume} {B782}},\
  \bibinfo {pages} {662} (\bibinfo {year} {2018})},\ \Eprint
  {http://arxiv.org/abs/1709.10054} {arXiv:1709.10054 [nucl-ex]} \BibitemShut
  {NoStop}%
\bibitem [{\citenamefont {de~Florian}\ \emph {et~al.}(2008)\citenamefont
  {de~Florian}, \citenamefont {Sassot}, \citenamefont {Stratmann},\ and\
  \citenamefont {Vogelsang}}]{deFlorian:2008mr}%
  \BibitemOpen
  \bibfield  {author} {\bibinfo {author} {\bibfnamefont {D.}~\bibnamefont
  {de~Florian}}, \bibinfo {author} {\bibfnamefont {R.}~\bibnamefont {Sassot}},
  \bibinfo {author} {\bibfnamefont {M.}~\bibnamefont {Stratmann}}, \ and\
  \bibinfo {author} {\bibfnamefont {W.}~\bibnamefont {Vogelsang}},\ }\href
  {\doibase 10.1103/PhysRevLett.101.072001} {\bibfield  {journal} {\bibinfo
  {journal} {Phys. Rev. Lett.}\ }\textbf {\bibinfo {volume} {101}},\ \bibinfo
  {pages} {072001} (\bibinfo {year} {2008})},\ \Eprint
  {http://arxiv.org/abs/0804.0422} {arXiv:0804.0422 [hep-ph]} \BibitemShut
  {NoStop}%
\bibitem [{\citenamefont {Leader}\ \emph {et~al.}(2010)\citenamefont {Leader},
  \citenamefont {Sidorov},\ and\ \citenamefont {Stamenov}}]{Leader:2010rb}%
  \BibitemOpen
  \bibfield  {author} {\bibinfo {author} {\bibfnamefont {E.}~\bibnamefont
  {Leader}}, \bibinfo {author} {\bibfnamefont {A.~V.}\ \bibnamefont {Sidorov}},
  \ and\ \bibinfo {author} {\bibfnamefont {D.~B.}\ \bibnamefont {Stamenov}},\
  }\href {\doibase 10.1103/PhysRevD.82.114018} {\bibfield  {journal} {\bibinfo
  {journal} {Phys. Rev.}\ }\textbf {\bibinfo {volume} {D82}},\ \bibinfo {pages}
  {114018} (\bibinfo {year} {2010})},\ \Eprint {http://arxiv.org/abs/1010.0574}
  {arXiv:1010.0574 [hep-ph]} \BibitemShut {NoStop}%
\bibitem [{\citenamefont {Mulders}\ and\ \citenamefont
  {Tangerman}(1996)}]{Mulders:1995dh}%
  \BibitemOpen
  \bibfield  {author} {\bibinfo {author} {\bibfnamefont {P.~J.}\ \bibnamefont
  {Mulders}}\ and\ \bibinfo {author} {\bibfnamefont {R.~D.}\ \bibnamefont
  {Tangerman}},\ }\href {\doibase 10.1016/0550-3213(95)00632-X} {\bibfield
  {journal} {\bibinfo  {journal} {Nucl. Phys.}\ }\textbf {\bibinfo {volume}
  {B461}},\ \bibinfo {pages} {197} (\bibinfo {year} {1996})},\ \Eprint
  {http://arxiv.org/abs/hep-ph/9510301} {hep-ph/9510301} \BibitemShut {NoStop}%
\bibitem [{\citenamefont {Bacchetta}\ \emph {et~al.}(2007)\citenamefont
  {Bacchetta}, \citenamefont {Diehl}, \citenamefont {Goeke}, \citenamefont
  {Metz}, \citenamefont {Mulders},\ and\ \citenamefont
  {Schlegel}}]{Bacchetta07}%
  \BibitemOpen
  \bibfield  {author} {\bibinfo {author} {\bibfnamefont {A.}~\bibnamefont
  {Bacchetta}}, \bibinfo {author} {\bibfnamefont {M.}~\bibnamefont {Diehl}},
  \bibinfo {author} {\bibfnamefont {K.}~\bibnamefont {Goeke}}, \bibinfo
  {author} {\bibfnamefont {A.}~\bibnamefont {Metz}}, \bibinfo {author}
  {\bibfnamefont {P.~J.}\ \bibnamefont {Mulders}}, \ and\ \bibinfo {author}
  {\bibfnamefont {M.}~\bibnamefont {Schlegel}},\ }\href {\doibase
  10.1088/1126-6708/2007/02/093} {\bibfield  {journal} {\bibinfo  {journal}
  {JHEP}\ }\textbf {\bibinfo {volume} {02}},\ \bibinfo {pages} {093} (\bibinfo
  {year} {2007})},\ \Eprint {http://arxiv.org/abs/hep-ph/0611265}
  {hep-ph/0611265} \BibitemShut {NoStop}%
\bibitem [{\citenamefont {Bacchetta}\ \emph {et~al.}(2008)\citenamefont
  {Bacchetta}, \citenamefont {Boer}, \citenamefont {Diehl},\ and\ \citenamefont
  {Mulders}}]{Bacchetta:2008xw}%
  \BibitemOpen
  \bibfield  {author} {\bibinfo {author} {\bibfnamefont {A.}~\bibnamefont
  {Bacchetta}}, \bibinfo {author} {\bibfnamefont {D.}~\bibnamefont {Boer}},
  \bibinfo {author} {\bibfnamefont {M.}~\bibnamefont {Diehl}}, \ and\ \bibinfo
  {author} {\bibfnamefont {P.~J.}\ \bibnamefont {Mulders}},\ }\href {\doibase
  10.1088/1126-6708/2008/08/023} {\bibfield  {journal} {\bibinfo  {journal}
  {JHEP}\ }\textbf {\bibinfo {volume} {08}},\ \bibinfo {pages} {023} (\bibinfo
  {year} {2008})},\ \Eprint {http://arxiv.org/abs/0803.0227} {arXiv:0803.0227
  [hep-ph]} \BibitemShut {NoStop}%
\bibitem [{\citenamefont {Bacchetta}\ \emph {et~al.}(2004)\citenamefont
  {Bacchetta}, \citenamefont {D'Alesio}, \citenamefont {Diehl},\ and\
  \citenamefont {Miller}}]{Bacchetta04}%
  \BibitemOpen
  \bibfield  {author} {\bibinfo {author} {\bibfnamefont {A.}~\bibnamefont
  {Bacchetta}}, \bibinfo {author} {\bibfnamefont {U.}~\bibnamefont {D'Alesio}},
  \bibinfo {author} {\bibfnamefont {M.}~\bibnamefont {Diehl}}, \ and\ \bibinfo
  {author} {\bibfnamefont {C.~A.}\ \bibnamefont {Miller}},\ }\href {\doibase
  10.1103/PhysRevD.70.117504} {\bibfield  {journal} {\bibinfo  {journal} {Phys.
  Rev.}\ }\textbf {\bibinfo {volume} {D70}},\ \bibinfo {pages} {117504}
  (\bibinfo {year} {2004})},\ \Eprint {http://arxiv.org/abs/hep-ph/0410050}
  {hep-ph/0410050} \BibitemShut {NoStop}%
\bibitem [{\citenamefont {Diehl}\ and\ \citenamefont
  {Sapeta}(2005)}]{Diehl:2005}%
  \BibitemOpen
  \bibfield  {author} {\bibinfo {author} {\bibfnamefont {M.}~\bibnamefont
  {Diehl}}\ and\ \bibinfo {author} {\bibfnamefont {S.}~\bibnamefont {Sapeta}},\
  }\href {\doibase 10.1140/epjc/s2005-02242-9} {\bibfield  {journal} {\bibinfo
  {journal} {Eur. Phys. J.}\ }\textbf {\bibinfo {volume} {C41}},\ \bibinfo
  {pages} {515} (\bibinfo {year} {2005})},\ \Eprint
  {http://arxiv.org/abs/hep-ph/0503023} {hep-ph/0503023} \BibitemShut {NoStop}%
\bibitem [{\citenamefont {Anselmino}\ \emph {et~al.}(2006)\citenamefont
  {Anselmino}, \citenamefont {Efremov}, \citenamefont {Kotzinian},\ and\
  \citenamefont {Parsamyan}}]{Anselmino:2006yc}%
  \BibitemOpen
  \bibfield  {author} {\bibinfo {author} {\bibfnamefont {M.}~\bibnamefont
  {Anselmino}}, \bibinfo {author} {\bibfnamefont {A.}~\bibnamefont {Efremov}},
  \bibinfo {author} {\bibfnamefont {A.}~\bibnamefont {Kotzinian}}, \ and\
  \bibinfo {author} {\bibfnamefont {B.}~\bibnamefont {Parsamyan}},\ }\href
  {\doibase 10.1103/PhysRevD.74.074015} {\bibfield  {journal} {\bibinfo
  {journal} {Phys. Rev.}\ }\textbf {\bibinfo {volume} {D74}},\ \bibinfo {pages}
  {074015} (\bibinfo {year} {2006})},\ \Eprint
  {http://arxiv.org/abs/hep-ph/0608048} {hep-ph/0608048} \BibitemShut {NoStop}%
\bibitem [{\citenamefont {Avakian}\ \emph {et~al.}(2007)\citenamefont
  {Avakian}, \citenamefont {Brodsky}, \citenamefont {Deur},\ and\ \citenamefont
  {Yuan}}]{Avakian:2007xa}%
  \BibitemOpen
  \bibfield  {author} {\bibinfo {author} {\bibfnamefont {H.}~\bibnamefont
  {Avakian}}, \bibinfo {author} {\bibfnamefont {S.~J.}\ \bibnamefont
  {Brodsky}}, \bibinfo {author} {\bibfnamefont {A.}~\bibnamefont {Deur}}, \
  and\ \bibinfo {author} {\bibfnamefont {F.}~\bibnamefont {Yuan}},\ }\href
  {\doibase 10.1103/PhysRevLett.99.082001} {\bibfield  {journal} {\bibinfo
  {journal} {Phys. Rev. Lett.}\ }\textbf {\bibinfo {volume} {99}},\ \bibinfo
  {pages} {082001} (\bibinfo {year} {2007})},\ \Eprint
  {http://arxiv.org/abs/0705.1553} {arXiv:0705.1553 [hep-ph]} \BibitemShut
  {NoStop}%
\bibitem [{\citenamefont {Musch}\ \emph {et~al.}(2011)\citenamefont {Musch},
  \citenamefont {H\"agler}, \citenamefont {Negele},\ and\ \citenamefont
  {Sch\"afer}}]{Musch:2010ka}%
  \BibitemOpen
  \bibfield  {author} {\bibinfo {author} {\bibfnamefont {B.~U.}\ \bibnamefont
  {Musch}}, \bibinfo {author} {\bibfnamefont {P.}~\bibnamefont {H\"agler}},
  \bibinfo {author} {\bibfnamefont {J.~W.}\ \bibnamefont {Negele}}, \ and\
  \bibinfo {author} {\bibfnamefont {A.}~\bibnamefont {Sch\"afer}},\ }\href
  {\doibase 10.1103/PhysRevD.83.094507} {\bibfield  {journal} {\bibinfo
  {journal} {Phys. Rev.}\ }\textbf {\bibinfo {volume} {D83}},\ \bibinfo {pages}
  {094507} (\bibinfo {year} {2011})},\ \Eprint {http://arxiv.org/abs/1011.1213}
  {arXiv:1011.1213 [hep-lat]} \BibitemShut {NoStop}%
\bibitem [{\citenamefont {Airapetian}\ \emph {et~al.}(2012)\citenamefont
  {Airapetian} \emph {et~al.}}]{Airapetian:2011wu}%
  \BibitemOpen
  \bibfield  {author} {\bibinfo {author} {\bibfnamefont {A.}~\bibnamefont
  {Airapetian}} \emph {et~al.} (\bibinfo {collaboration} {The HERMES
  Collaboration}),\ }\href {\doibase 10.1140/epjc/s10052-012-1921-5} {\bibfield
   {journal} {\bibinfo  {journal} {Eur. Phys. J.}\ }\textbf {\bibinfo {volume}
  {C72}},\ \bibinfo {pages} {1921} (\bibinfo {year} {2012})},\ \Eprint
  {http://arxiv.org/abs/1112.5584} {arXiv:1112.5584 [hep-ex]} \BibitemShut
  {NoStop}%
\bibitem [{\citenamefont {Wandzura}\ and\ \citenamefont
  {Wilczek}(1977)}]{Wandzura:1977qf}%
  \BibitemOpen
  \bibfield  {author} {\bibinfo {author} {\bibfnamefont {S.}~\bibnamefont
  {Wandzura}}\ and\ \bibinfo {author} {\bibfnamefont {F.}~\bibnamefont
  {Wilczek}},\ }\href {\doibase 10.1016/0370-2693(77)90700-6} {\bibfield
  {journal} {\bibinfo  {journal} {Phys. Lett.}\ }\textbf {\bibinfo {volume}
  {B72}},\ \bibinfo {pages} {195} (\bibinfo {year} {1977})}\BibitemShut
  {NoStop}%
\bibitem [{\citenamefont {Oganessyan}\ \emph {et~al.}(2002)\citenamefont
  {Oganessyan}, \citenamefont {Mulders},\ and\ \citenamefont
  {De~Sanctis}}]{Oganessyan:2002pc}%
  \BibitemOpen
  \bibfield  {author} {\bibinfo {author} {\bibfnamefont {K.~A.}\ \bibnamefont
  {Oganessyan}}, \bibinfo {author} {\bibfnamefont {P.~J.}\ \bibnamefont
  {Mulders}}, \ and\ \bibinfo {author} {\bibfnamefont {E.}~\bibnamefont
  {De~Sanctis}},\ }\href {\doibase 10.1016/S0370-2693(02)01532-0} {\bibfield
  {journal} {\bibinfo  {journal} {Phys. Lett.}\ }\textbf {\bibinfo {volume}
  {B532}},\ \bibinfo {pages} {87} (\bibinfo {year} {2002})},\ \Eprint
  {http://arxiv.org/abs/hep-ph/0201061} {arXiv:hep-ph/0201061 [hep-ph]}
  \BibitemShut {NoStop}%
\bibitem [{\citenamefont {Boer}\ and\ \citenamefont
  {Mulders}(1998)}]{Boer:1997nt}%
  \BibitemOpen
  \bibfield  {author} {\bibinfo {author} {\bibfnamefont {D.}~\bibnamefont
  {Boer}}\ and\ \bibinfo {author} {\bibfnamefont {P.~J.}\ \bibnamefont
  {Mulders}},\ }\href {\doibase 10.1103/PhysRevD.57.5780} {\bibfield  {journal}
  {\bibinfo  {journal} {Phys. Rev.}\ }\textbf {\bibinfo {volume} {D57}},\
  \bibinfo {pages} {5780} (\bibinfo {year} {1998})},\ \Eprint
  {http://arxiv.org/abs/hep-ph/9711485} {arXiv:hep-ph/9711485 [hep-ph]}
  \BibitemShut {NoStop}%
\bibitem [{\citenamefont {Ravndal}(1973)}]{Ravndal73}%
  \BibitemOpen
  \bibfield  {author} {\bibinfo {author} {\bibfnamefont {F.}~\bibnamefont
  {Ravndal}},\ }\href {\doibase 10.1016/0370-2693(73)90445-0} {\bibfield
  {journal} {\bibinfo  {journal} {Physics Letters}\ }\textbf {\bibinfo {volume}
  {B43}},\ \bibinfo {pages} {301 } (\bibinfo {year} {1973})}\BibitemShut
  {NoStop}%
\bibitem [{\citenamefont {Kingsley}(1974)}]{Kingsley74}%
  \BibitemOpen
  \bibfield  {author} {\bibinfo {author} {\bibfnamefont {R.~L.}\ \bibnamefont
  {Kingsley}},\ }\href {\doibase 10.1103/PhysRevD.10.1580} {\bibfield
  {journal} {\bibinfo  {journal} {Phys. Rev.}\ }\textbf {\bibinfo {volume}
  {D10}},\ \bibinfo {pages} {1580} (\bibinfo {year} {1974})}\BibitemShut
  {NoStop}%
\bibitem [{\citenamefont {Cahn}(1978)}]{NCahn1978269}%
  \BibitemOpen
  \bibfield  {author} {\bibinfo {author} {\bibfnamefont {R.~N.}\ \bibnamefont
  {Cahn}},\ }\href {\doibase DOI: 10.1016/0370-2693(78)90020-5} {\bibfield
  {journal} {\bibinfo  {journal} {Phys. Lett.}\ }\textbf {\bibinfo {volume}
  {B78}},\ \bibinfo {pages} {269 } (\bibinfo {year} {1978})}\BibitemShut
  {NoStop}%
\bibitem [{\citenamefont {Airapetian}\ \emph
  {et~al.}(2013{\natexlab{a}})\citenamefont {Airapetian} \emph
  {et~al.}}]{Airapetian:2012yg}%
  \BibitemOpen
  \bibfield  {author} {\bibinfo {author} {\bibfnamefont {A.}~\bibnamefont
  {Airapetian}} \emph {et~al.} (\bibinfo {collaboration} {HERMES
  Collaboration}),\ }\href {\doibase 10.1103/PhysRevD.87.012010} {\bibfield
  {journal} {\bibinfo  {journal} {Phys. Rev.}\ }\textbf {\bibinfo {volume}
  {D87}},\ \bibinfo {pages} {012010} (\bibinfo {year} {2013}{\natexlab{a}})},\
  \Eprint {http://arxiv.org/abs/1204.4161} {arXiv:1204.4161 [hep-ex]}
  \BibitemShut {NoStop}%
\bibitem [{\citenamefont {Adolph}\ \emph {et~al.}(2014)\citenamefont {Adolph}
  \emph {et~al.}}]{Adolph:2014pwc}%
  \BibitemOpen
  \bibfield  {author} {\bibinfo {author} {\bibfnamefont {C.}~\bibnamefont
  {Adolph}} \emph {et~al.} (\bibinfo {collaboration} {COMPASS Collaboration}),\
  }\href {\doibase 10.1016/j.nuclphysb.2014.07.019} {\bibfield  {journal}
  {\bibinfo  {journal} {Nucl. Phys.}\ }\textbf {\bibinfo {volume} {B886}},\
  \bibinfo {pages} {1046} (\bibinfo {year} {2014})},\ \Eprint
  {http://arxiv.org/abs/1401.6284} {arXiv:1401.6284 [hep-ex]} \BibitemShut
  {NoStop}%
\bibitem [{\citenamefont {Yan}\ \emph {et~al.}(2017)\citenamefont {Yan} \emph
  {et~al.}}]{Yan:2016ods}%
  \BibitemOpen
  \bibfield  {author} {\bibinfo {author} {\bibfnamefont {X.}~\bibnamefont
  {Yan}} \emph {et~al.} (\bibinfo {collaboration} {Jefferson Lab Hall A
  Collaboration}),\ }\href {\doibase 10.1103/PhysRevC.95.035209} {\bibfield
  {journal} {\bibinfo  {journal} {Phys. Rev.}\ }\textbf {\bibinfo {volume}
  {C95}},\ \bibinfo {pages} {035209} (\bibinfo {year} {2017})},\ \Eprint
  {http://arxiv.org/abs/1610.02350} {arXiv:1610.02350 [nucl-ex]} \BibitemShut
  {NoStop}%
\bibitem [{\citenamefont {Frankfurt}\ \emph {et~al.}(1989)\citenamefont
  {Frankfurt}, \citenamefont {Strikman}, \citenamefont {Mankiewicz},
  \citenamefont {Sch{\"a}fer}, \citenamefont {Rondio}, \citenamefont
  {Sandacz},\ and\ \citenamefont {Papavassiliou}}]{Frankfurt:1989wq}%
  \BibitemOpen
  \bibfield  {author} {\bibinfo {author} {\bibfnamefont {L.~L.}\ \bibnamefont
  {Frankfurt}}, \bibinfo {author} {\bibfnamefont {M.~I.}\ \bibnamefont
  {Strikman}}, \bibinfo {author} {\bibfnamefont {L.}~\bibnamefont
  {Mankiewicz}}, \bibinfo {author} {\bibfnamefont {A.}~\bibnamefont
  {Sch{\"a}fer}}, \bibinfo {author} {\bibfnamefont {E.}~\bibnamefont {Rondio}},
  \bibinfo {author} {\bibfnamefont {A.}~\bibnamefont {Sandacz}}, \ and\
  \bibinfo {author} {\bibfnamefont {V.}~\bibnamefont {Papavassiliou}},\ }\href
  {\doibase 10.1016/0370-2693(89)91668-7} {\bibfield  {journal} {\bibinfo
  {journal} {Phys. Lett.}\ }\textbf {\bibinfo {volume} {B230}},\ \bibinfo
  {pages} {141} (\bibinfo {year} {1989})}\BibitemShut {NoStop}%
\bibitem [{\citenamefont {Ackerstaff}\ \emph {et~al.}(1998)\citenamefont
  {Ackerstaff} \emph {et~al.}}]{Ackerstaff:1998av}%
  \BibitemOpen
  \bibfield  {author} {\bibinfo {author} {\bibfnamefont {K.}~\bibnamefont
  {Ackerstaff}} \emph {et~al.} (\bibinfo {collaboration} {The HERMES
  Collaboration}),\ }\href {\doibase 10.1016/S0168-9002(98)00769-4} {\bibfield
  {journal} {\bibinfo  {journal} {Nucl. Instrum. Meth.}\ }\textbf {\bibinfo
  {volume} {A417}},\ \bibinfo {pages} {230} (\bibinfo {year} {1998})},\ \Eprint
  {http://arxiv.org/abs/hep-ex/9806008} {hep-ex/9806008} \BibitemShut {NoStop}%
\bibitem [{\citenamefont {Airapetian}\ \emph {et~al.}(2007)\citenamefont
  {Airapetian} \emph {et~al.}}]{Airapetian:2007mh}%
  \BibitemOpen
  \bibfield  {author} {\bibinfo {author} {\bibfnamefont {A.}~\bibnamefont
  {Airapetian}} \emph {et~al.} (\bibinfo {collaboration} {HERMES
  Collaboration}),\ }\href {\doibase 10.1103/PhysRevD.75.012007} {\bibfield
  {journal} {\bibinfo  {journal} {Phys. Rev.}\ }\textbf {\bibinfo {volume}
  {D75}},\ \bibinfo {pages} {012007} (\bibinfo {year} {2007})},\ \Eprint
  {http://arxiv.org/abs/hep-ex/0609039} {hep-ex/0609039} \BibitemShut {NoStop}%
\bibitem [{\citenamefont {Airapetian}\ \emph
  {et~al.}(2005{\natexlab{b}})\citenamefont {Airapetian} \emph
  {et~al.}}]{Airapetian:2005cb}%
  \BibitemOpen
  \bibfield  {author} {\bibinfo {author} {\bibfnamefont {A.}~\bibnamefont
  {Airapetian}} \emph {et~al.} (\bibinfo {collaboration} {HERMES
  Collaboration}),\ }\href {\doibase 10.1103/PhysRevLett.95.242001} {\bibfield
  {journal} {\bibinfo  {journal} {Phys. Rev. Lett.}\ }\textbf {\bibinfo
  {volume} {95}},\ \bibinfo {pages} {242001} (\bibinfo {year}
  {2005}{\natexlab{b}})},\ \Eprint {http://arxiv.org/abs/hep-ex/0506018}
  {hep-ex/0506018 [hep-ex]} \BibitemShut {NoStop}%
\bibitem [{\citenamefont {Abe}\ \emph {et~al.}(1999)\citenamefont {Abe} \emph
  {et~al.}}]{collaboration-1999-452}%
  \BibitemOpen
  \bibfield  {author} {\bibinfo {author} {\bibfnamefont {K.}~\bibnamefont
  {Abe}} \emph {et~al.} (\bibinfo {collaboration} {E143 Collaboration}),\
  }\href {doi:10.1016/S0370-2693(99)00244-0} {\bibfield  {journal} {\bibinfo
  {journal} {Phys. Lett.}\ }\textbf {\bibinfo {volume} {B452}},\ \bibinfo
  {pages} {194} (\bibinfo {year} {1999})},\ \Eprint
  {http://arxiv.org/abs/hep-ex/9808028} {hep-ex/9808028} \BibitemShut {NoStop}%
\bibitem [{\citenamefont {Rondon}(1999)}]{physRevC.60.035201}%
  \BibitemOpen
  \bibfield  {author} {\bibinfo {author} {\bibfnamefont {O.~A.}\ \bibnamefont
  {Rondon}},\ }\href {\doibase 10.1103/PhysRevC.60.035201} {\bibfield
  {journal} {\bibinfo  {journal} {Phys. Rev.}\ }\textbf {\bibinfo {volume}
  {C60}},\ \bibinfo {pages} {035201} (\bibinfo {year} {1999})}\BibitemShut
  {NoStop}%
\bibitem [{\citenamefont {Akopov}\ \emph {et~al.}(2002)\citenamefont {Akopov}
  \emph {et~al.}}]{Akopov:2000qi}%
  \BibitemOpen
  \bibfield  {author} {\bibinfo {author} {\bibfnamefont {N.}~\bibnamefont
  {Akopov}} \emph {et~al.},\ }\href {\doibase 10.1016/S0168-9002(01)00932-9}
  {\bibfield  {journal} {\bibinfo  {journal} {Nucl. Instrum. Meth.}\ }\textbf
  {\bibinfo {volume} {A479}},\ \bibinfo {pages} {511} (\bibinfo {year}
  {2002})},\ \Eprint {http://arxiv.org/abs/physics/0104033} {physics/0104033}
  \BibitemShut {NoStop}%
\bibitem [{\citenamefont {Airapetian}\ \emph
  {et~al.}(2013{\natexlab{b}})\citenamefont {Airapetian} \emph
  {et~al.}}]{Airapetian:2012ki}%
  \BibitemOpen
  \bibfield  {author} {\bibinfo {author} {\bibfnamefont {A.}~\bibnamefont
  {Airapetian}} \emph {et~al.} (\bibinfo {collaboration} {HERMES
  Collaboration}),\ }\href {\doibase 10.1103/PhysRevD.87.074029} {\bibfield
  {journal} {\bibinfo  {journal} {Phys. Rev.}\ }\textbf {\bibinfo {volume}
  {D87}},\ \bibinfo {pages} {074029} (\bibinfo {year} {2013}{\natexlab{b}})},\
  \Eprint {http://arxiv.org/abs/1212.5407} {arXiv:1212.5407 [hep-ex]}
  \BibitemShut {NoStop}%
\bibitem [{\citenamefont {Airapetian}\ \emph
  {et~al.}(2005{\natexlab{c}})\citenamefont {Airapetian} \emph
  {et~al.}}]{Airapetian:2005jc}%
  \BibitemOpen
  \bibfield  {author} {\bibinfo {author} {\bibfnamefont {A.}~\bibnamefont
  {Airapetian}} \emph {et~al.} (\bibinfo {collaboration} {HERMES
  Collaboration}),\ }\href {\doibase 10.1016/j.physletb.2005.06.067} {\bibfield
   {journal} {\bibinfo  {journal} {Phys. Lett.}\ }\textbf {\bibinfo {volume}
  {B622}},\ \bibinfo {pages} {14} (\bibinfo {year} {2005}{\natexlab{c}})},\
  \Eprint {http://arxiv.org/abs/hep-ex/0505042} {hep-ex/0505042} \BibitemShut
  {NoStop}%
\bibitem [{\citenamefont {Alekseev}\ \emph
  {et~al.}(2010{\natexlab{b}})\citenamefont {Alekseev} \emph
  {et~al.}}]{Alekseev:2010dm}%
  \BibitemOpen
  \bibfield  {author} {\bibinfo {author} {\bibfnamefont {M.}~\bibnamefont
  {Alekseev}} \emph {et~al.} (\bibinfo {collaboration} {COMPASS
  Collaboration}),\ }\href {\doibase 10.1140/epjc/s10052-010-1461-9} {\bibfield
   {journal} {\bibinfo  {journal} {Eur. Phys. J.}\ }\textbf {\bibinfo {volume}
  {C70}},\ \bibinfo {pages} {39} (\bibinfo {year} {2010}{\natexlab{b}})},\
  \Eprint {http://arxiv.org/abs/1007.1562} {arXiv:1007.1562 [hep-ex]}
  \BibitemShut {NoStop}%
\bibitem [{\citenamefont {Adolph}\ \emph {et~al.}(2016)\citenamefont {Adolph}
  \emph {et~al.}}]{Adolph:2016vou}%
  \BibitemOpen
  \bibfield  {author} {\bibinfo {author} {\bibfnamefont {C.}~\bibnamefont
  {Adolph}} \emph {et~al.} (\bibinfo {collaboration} {COMPASS Collaboration}),\
  }\href@noop {} {\  (\bibinfo {year} {2016})},\ \Eprint
  {http://arxiv.org/abs/1609.06062} {arXiv:1609.06062 [hep-ex]} \BibitemShut
  {NoStop}%
\bibitem [{asy()}]{asymData}%
  \BibitemOpen
  \href@noop {} {}\bibinfo {note} {Supplemental material (this publication);
  mail-to: \href{mailto:management@hermes.desy.de}{management@hermes.desy.de};
  HERMES web:
  \href{http://www-hermes.desy.de/notes/pub/publications.html}{http://www-hermes.desy.de/notes/pub/publications.html}}\BibitemShut
  {NoStop}%
\end{thebibliography}%

\end{document}